\documentclass[10pt,a4paper]{article}
 \pdfoutput=1
 \usepackage[dvipsnames]{xcolor}
\usepackage{amsmath,fourier,amssymb,amsthm,graphicx,epstopdf,tikz,
color,cases}
\usetikzlibrary{shapes,snakes}
\usepackage{chngcntr}
\counterwithout{figure}{section}
\usepackage[T1]{fontenc}
\allowdisplaybreaks
\numberwithin{equation}{section}

 \newtheorem{innercustomthm}{Theorem}
 \newenvironment{customthm}[1]
   {\renewcommand\theinnercustomthm{#1}\innercustomthm}
   {\endinnercustomthm}

 \newtheorem{innercustomconj}{Conjecture}
 \newenvironment{customconj}[1]
   {\renewcommand\theinnercustomconj{#1}\innercustomconj}
    {\endinnercustomconj}

 \theoremstyle{plain}
 \newtheorem {hypo}{\bf\hspace{-\parindent}Hypothesis}[section]

 \newtheorem {prop}[hypo]{Proposition}
 \newtheorem {lemma}[hypo]{Lemma}
 
 \newtheorem {defin}[hypo]{Definition}
 \newtheorem {cor}[hypo]{Corollary}

 \theoremstyle{remark}
 \newtheorem {rmk}[hypo]{Remark}
  
 \newcommand{\pf}{\begin{bpf}}

 \newcommand{\pfms}{\begin{bpfms}}
 \newcommand{\epf}{\end{bpf}\hfill$\square$\vspace{0.1cm}}
 \newcommand{\epfms}{\end{bpfms}\hfill$\square$\\ }
 \newcommand\ben{\begin{equation*}}
 \newcommand\ebn{\end{equation*}}
 \newcommand\beq{\begin{equation}}
 \newcommand\eeq{\end{equation}}
 \newcommand\ds{\displaystyle}
  \newcommand\lb{\left(}
  \newcommand\rb{\right)} 
  
   \newcommand\Cb{\mathbb{C}} 
   
   \newcommand\Pb{\mathbb{P}}

\begin{document}
\LARGE
\noindent
\textbf{Irregular conformal blocks  and connection formulae \\ for Painlevé V functions}
\normalsize
 \vspace{1cm}\\
  \noindent\textit{
  O. Lisovyy$\,^{a,}$\footnote{lisovyi@lmpt.univ-tours.fr},
  H. Nagoya$\,^{b,}$\footnote{nagoya@se.kanazawa-u.ac.jp},
  J. Roussillon$\,^{a,}$\footnote{Julien.Roussillon@lmpt.univ-tours.fr}}
  \vspace{0.2cm}\\
  $^a$ Institut Denis-Poisson, Universit\'e de Tours, Université d'Orléans, CNRS, Parc de Grandmont,
   37200 Tours, France\vspace{0.1cm}\\
  $^b$ School of Mathematics and Physics, Kanazawa University, Kanazawa, Ishikawa 920-1192, Japan

   \begin{abstract}
   We prove a Fredholm determinant and short-distance series representation of the Painlevé V tau function~$\tau\lb t\rb$ associated to generic monodromy data. Using a relation of $\tau\lb t\rb$ to two different types of irregular $c=1$ Virasoro conformal blocks and the confluence from Painlevé VI equation, connection formulas between the parameters of asymptotic expansions at $0$ and $i\infty$ are conjectured. Explicit evaluations of the connection constants relating the tau function asymptotics as $t\to0,+\infty,i\infty$ are obtained. 
   We also show that irregular conformal blocks of rank 1, for arbitrary central charge, are obtained as
   confluent limits of the regular conformal blocks.
   \end{abstract}
 
 \section{Introduction and description of results}
 This paper is concerned with two related classes of special functions: Virasoro conformal blocks (CBs) and Painlev\'e transcendents. The realm of conformal blocks is 2D CFT \cite{BPZ,DFMS} (representation theory of the Virasoro algebra), whereas Painlev\'e functions come from the theory of monodromy preserving deformations of linear ODEs with rational coefficients \cite{JMU,FIKN}.
 
 \subsection{Regular conformal blocks} The (regular) $4$-point spherical CB is often represented  in the  literature by a trivalent graph encoding the expectation value of a composition of two primary vertex operators,
 \ben
 \begin{tikzpicture}[baseline,yshift=-0.3cm,scale=0.8]
 \draw [thick] (-1,0) -- (2,0);
 \draw [thick] (0,0) -- (0,1);
 \draw [thick] (1,0) -- (1,1); 
 \draw (-0.7,0) node[above] {\scriptsize $\theta_\infty$};
 \draw (0.5,0) node[above] {\scriptsize $\sigma$};
 \draw (1.7,0) node[above] {\scriptsize $\theta_0$};
 \draw (0,1) node[left] {\scriptsize $\theta_1$};
 \draw (1,1) node[right] {\scriptsize $\theta_t$};
 \end{tikzpicture}
  \lb t\rb \equiv \mathcal F\lb\substack{\theta_{1}\;\quad 
  \theta_{t}\\ \sigma \\ \theta_{\infty}\quad \theta_0};t\rb: =\left\langle \Delta_{\infty}\right|V_{\Delta_{\infty},\Delta}^{\Delta_1}\lb 1\rb V_{\Delta,\Delta_0}^{\Delta_t}\lb t\rb\left|\Delta_0\right\rangle.
 \ebn 
 This function depends on the Virasoro central charge $c$,  five conformal dimensions $\Delta\lb \sigma\rb=\frac{c-1}{24}+\sigma^2$ attached to edges and labeling highest weight modules, as well as the anharmonic ratio $t$ of $4$ points on $\Cb\Pb^1$. The vertices of the graph represent chiral vertex operators.

 As a function of $t$, the regular CB is given by an expansion of the form $\mathcal F\lb t\rb= t^{\Delta-\Delta_0-\Delta_t}\left[1+\sum_{k=1}^{\infty}\mathcal F_k t^k\right]$. The series inside the brackets is believed to be convergent inside the unit disk $|t|<1$ and to continue analytically to the universal covering of $\Cb\backslash \left\{0,1\right\}$ \cite{Zam87,HJP}. Under this assumption,
 CB becomes naturally defined for ${t\in\Cb\backslash\lb\lb -\infty,0]\cup[1,\infty\rb\rb }$. Its analytic properties resemble those of the Gauss hypergeometric function, recovered in the classical limit
 \beq
 \label{classlim2f1}
 \lim_{c\to\infty}\mathcal F\lb\substack{\theta_{1}\;\quad 
   \theta_{t}\\ \sigma \\ \theta_{\infty}\quad \theta_0};t\rb=t^{\Delta-\Delta_0-\Delta_t}
 {}_2F_1\left(\substack{\Delta_t-\Delta_0+\Delta,\Delta_1-\Delta_{\infty}+\Delta\\ 2\Delta};t\right).
 \eeq
 CB is expected to be analytic in the external momenta $\theta_{0,t,1,\infty}$ and meromorphic in the internal momentum $\sigma$, with the only possible poles located at $\pm\sigma_{m,n}=\frac{1}{2}\lb m\beta-n\beta^{-1}\rb$, where $m,n\in\mathbb Z_{\ge0}$ and $\beta$ appears in the Liouville-type parameterization of the central charge:  $c=1-6Q^2$, $Q=\beta-\beta^{-1}$.
 
 Besides their manifest invariance with respect to the sign flips of  momenta,
 regular CBs possess the following discrete symmetries:
   \begin{alignat*}{3}
   \mathcal F\lb\substack{\theta_{1}\;\quad 
      \theta_{t}\\ \sigma \\ \theta_{\infty}\quad \theta_0};t\rb=&\,t^{\frac{\lb\theta_1+\theta_{\infty}\rb^2-
                 \lb\theta_t+\theta_0\rb^2}{2}}\lb 1-t\rb^{\frac{\lb\theta_0+\theta_{\infty}\rb^2-
            \lb\theta_t+\theta_1\rb^2}{2}} \mathcal F\lb\substack{\theta_{1}-\delta\;\quad 
                \theta_{t}-\delta\\ \sigma \\ \theta_{\infty}-\delta\quad \theta_0-\delta};t\rb,   \qquad && \text{(Regge-Okamoto symmetry)}\\
       =&\, e^{\pm i\pi \lb \Delta-\Delta_0-\Delta_t\rb} \lb 1-t\rb^{-2\Delta_t} \mathcal F\lb\substack{\theta_{\infty}\quad 
                \theta_{t}\\ \sigma \\ \theta_{1}\;\quad \theta_0};\frac{t}{t-1}\rb,\quad \Im t\gtrless 0,\qquad&&\text{(braiding transformation)}
   \end{alignat*}
  where  $\delta=\frac12\lb\theta_{0}+\theta_t+\theta_1+\theta_{\infty}\rb$.  Three flips $s_{\nu}:\theta_{\nu}\mapsto -\theta_{\nu}$ ($\nu=0,t,1$)  and Regge-Okamoto transformation~$s_\delta$ generate an action of the Weyl group of type $D_4$ on the external momenta. The 4th flip $s_{\infty}:\theta_{\infty}\mapsto-\theta_{\infty}$ may be obtained as  $s_{\infty}=s_{\delta}s_ts_0s_{\delta}s_1
  s_{\delta}s_ts_0s_{\delta}$. The compositions such as 
  $s_c=s_{\infty}s_0s_{\delta}s_1 s_{t}s_{\delta}$ and $s_r=s_{\infty}s_1 s_{\delta}s_t s_{0}s_{\delta}$ lead to more familiar symmetries
  \begin{alignat*}{3}
  \qquad\qquad\quad\mathcal F\lb\substack{\theta_{1}\;\quad 
     \theta_{t}\\ \sigma \\ \theta_{\infty}\quad \theta_0};t\rb=&\,
      t^{\Delta_{\infty}+\Delta_1-\Delta_t-\Delta_0}\mathcal F\lb\substack{\theta_{t}\;\quad 
    \theta_{1}\\ \sigma \\ \theta_{0}\quad \theta_{\infty}};t\rb=
    &&\text{(exchange of columns)} \\
    =&\,\lb 1-t\rb^{\Delta_0-\Delta_t-\Delta_1+\Delta_{\infty}}\mathcal F\lb\substack{\theta_{\infty}\quad \theta_0\\ \sigma \\ \theta_{1}\;\quad     \theta_{t} };t\rb.\qquad \qquad&&\text{(exchange of rows)} 
  \end{alignat*}
  For reader's convenience, we record the action of generators of $W\lb D_4\rb$ and their special combinations in the following table:
  \begin{center}
  \begin{tabular}{c||c|c|c|c}
  & $\theta_0$ & $\theta_t$ & $\theta_1$ & $\theta_{\infty} $ \\ \hline\hline
  $s_0$ & $-\theta_0$ & $\theta_t$ & $\theta_1$ & $\theta_{\infty} $ \\ \hline
  $s_t$ & $\theta_0$ & $-\theta_t$ & $\theta_1$ & $\theta_{\infty} $ \\ \hline
  $s_1$ & $\theta_0$ & $\theta_t$ & $-\theta_1$ & $\theta_{\infty} $ \\ \hline
  $s_{\delta}$ & $\theta_0-\delta$ & $\theta_t-\delta$ & $\theta_1-\delta$ & $\theta_{\infty}-\delta$ \\ \hline\hline
  $s_{\infty}$ & $\theta_0$ & $\theta_t$ & $\theta_1$ & $-\theta_{\infty} $ \\ \hline
  $s_c$ & $\theta_{\infty}$ & $\theta_1$ & $\theta_t$ & $\theta_0 $ \\ \hline
  $s_r$ & $\theta_t$ & $\theta_0$ & $\theta_{\infty}$ & $\theta_1$
  \end{tabular}
  \end{center}
  
  Additional nontrivial transformations correspond to crossing symmetry. They relate CBs associated to 3~different ways of splitting 4 points on $\Cb\Pb^1$ into 2 pairs which are called $s$-, $t$- and $u$-channel. The cross-ratio arguments of CBs in these channels are chosen from
 $\left\{t,\frac{t}{t-1}\right\}$, $\left\{1-t,\frac{t-1}{t}\right\}$ and $\left\{\frac1t,\frac{1}{1-t}\right\}$, respectively. All crossing transformations are generated by the above discrete symmetries and one basic relation (diagonal flip)
 \ben
 \qquad\mathcal F\lb\substack{\theta_{1}\;\quad 
     \theta_{t}\\ \sigma \\ \theta_{\infty}\quad \theta_0};t\rb=\int_{\mathbb R^+}
 F\left[\substack{\theta_1\;\;\;\theta_t\vspace{0.1cm}\\ \theta_{\infty}\;\;\theta_0};\substack{\rho\vspace{0.15cm} \\  \sigma}\right]         
     \mathcal F\lb\substack{\theta_{0}\;\quad 
          \theta_{t}\\ \rho \\ \theta_{\infty}\quad \theta_1};1-t\rb
          d\rho,\qquad\quad \text{(fusion relation)}
 \ebn
 with $c\notin(-\infty,1]$. The fusion kernel $F$ in the above formula is independent of $t$ and can be explicitly expressed in terms of Faddeev's quantum dilogarithms \cite{pt1,pt2}. The integrand is regular for $\rho\in\mathbb R^+$; in particular, it has no singularity at $\rho=0$ since the relevant pole of $\mathcal F$ is compensated by the zero of $F$. The condition on the central charge ensures that the Ponsot-Teschner kernel is well-defined. 
 CBs in different channels thus provide different bases of the same
 infinite-dimensional space $\bigoplus\limits_{\sigma\in\mathbb R^+}\mathbb C\,
 \mathcal F\lb\substack{\theta_{1}\;\quad 
      \theta_{t}\\ \sigma \\ \theta_{\infty}\quad \theta_0};t\rb$. This is analogous to pairs of fundamental solutions associated to different singular points of the hypergeometric equation.

 The series representation for conformal blocks was made completely explicit by the discovery of the AGT relation \cite{AGT} between the 2D CFTs and 4D SUSY gauge theories. Denoting by $\mathbb Y$ the set of Young diagrams, the 4-point CB is expressed as \cite{AFLT,Nekrasov}
 \beq\label{CBseries}
 \mathcal F\lb\substack{\theta_{1}\;\quad 
      \theta_{t}\\ \sigma \\ \theta_{\infty}\quad \theta_0};t\rb=
      t^{\Delta-\Delta_0-\Delta_t}\lb 1-t\rb^{2\lb\theta_t+\frac Q2\rb
      \lb\theta_1+\frac Q2\rb}\sum_{\lambda,\mu\in\mathbb Y} \mathcal F_{\lambda,\mu}\lb\substack{\theta_{1}\;\quad 
            \theta_{t}\\ \sigma \\ \theta_{\infty}\quad \theta_0}\rb t^{|\lambda|+|\mu|},
 \eeq
 where $|\lambda|$ denotes the number of boxes in the diagram $\lambda\in\mathbb Y$.
 In order to write the coefficients $\mathcal F_{\lambda,\mu}$ explicitly, denote by
 $a_{\lambda}\lb \square\rb$ and $l_{\lambda}\lb \square\rb$ the arm-length and leg-length of the box $\square$ in $\lambda$. Note that the box in question does not necessarily belong to $\lambda$, therefore $a_{\lambda}\lb \square\rb$ and/or $l_{\lambda}\lb \square\rb$ may be negative. For any for  $\theta\in\mathbb C$ and $\lambda,\mu\in\mathbb Y$, introduce Nekrasov functions
 \beq\label{nekrofun}
 Z_{\lambda,\mu}\lb \theta\rb=Z_{\mu,\lambda}\lb -\theta\rb=\prod_{\square\in\lambda}
 \Bigl( \beta^{-1}\lb a_{\lambda}\lb\square\rb+{\tfrac12}\rb+\beta \lb l_{\mu}\lb \square\rb +\tfrac12\rb+\theta\Bigr)\prod_{\square\in\mu}
  \Bigl( \beta^{-1}\lb a_{\mu}\lb\square\rb+{\tfrac12}\rb+\beta \lb l_{\lambda}\lb \square\rb +\tfrac12\rb-\theta\Bigr),
 \eeq
 The expansion coefficients $\mathcal F_{\lambda,\mu}$ can then be represented by the factorized expressions 
 \beq
 \label{rCBcoef}
 \mathcal F_{\lambda,\mu}\lb\substack{\theta_{1}\;\quad 
             \theta_{t}\\ \sigma \\ \theta_{\infty}\quad \theta_0}\rb=\frac{\prod_{\epsilon=\pm}
 Z_{\emptyset,\lambda}\lb\epsilon\theta_0-\theta_t-\sigma\rb
 Z_{\emptyset,\mu}\lb\epsilon\theta_0-\theta_t+\sigma\rb         Z_{\lambda,\emptyset}\lb\epsilon\theta_{\infty}+\theta_1
 +\sigma\rb  Z_{\mu,\emptyset}\lb\epsilon\theta_{\infty}+\theta_1-\sigma\rb
             }{Z_{\lambda,\lambda}\lb \tfrac Q2\rb  
             Z_{\mu,\mu}\lb \tfrac Q2\rb Z_{\lambda,\mu}\lb \tfrac Q2+2\sigma\rb Z_{\mu,\lambda}\lb \tfrac Q2-2\sigma\rb}.
 \eeq
  The invariance of CBs with respect to the sign flips of $\theta_t$, $\theta_1$ or $Q$ as well as the braiding transformation are no longer manifest in the series representation (\ref{CBseries}). On the other hand, the Regge-Okamoto symmetry is seen rather easily.
 
 \subsection{Confluent conformal blocks of the 1st kind}
 The AGT relation has also triggered the study of irregular conformal blocks \cite{G09,BMT,GT}. One class of them, relevant for the present work, corresponds to decoupling of one of the four matter multiplets on the gauge side. It is described by the scaling limit 
 \beq\label{conflimit}
 \theta_1=\frac{\Lambda+\theta_*}{2},\qquad 
 \theta_{\infty}=\frac{\Lambda-\theta_*}{2}, \qquad
 t\mapsto\frac{t}{\Lambda},\qquad \Lambda\to\infty.
 \eeq
 Calculating this limit termwise in \eqref{CBseries}, it may be easily checked that the number of $Z$-factors in the numerator of $\mathcal F_{\lambda,\mu}$ reduces from eight to six. Namely, if we consider
 \beq\label{confCB1series}
 \mathcal B\lb \theta_*;\sigma;\substack{\theta_t \\ \theta_0};t\rb:=\lim_{\Lambda\to\infty}
 \Lambda^{\Delta-\Delta_0-\Delta_t}
 \mathcal F\lb\substack{\frac{\Lambda+\theta_*}{2}\quad 
              \theta_{t}\\ \quad\sigma \\ \frac{\Lambda-\theta_*}{2}\quad \theta_0};\frac{t}{\Lambda}\rb=
 t^{\Delta-\Delta_0-\Delta_t}e^{-\lb\theta_t+\frac Q2\rb t}\sum_{\lambda,\mu\in\mathbb Y}
 \mathcal B_{\lambda,\mu}\lb \theta_*;\sigma;\substack{\theta_t \\ \theta_0}\rb
 t^{|\lambda|+|\mu|},
 \eeq
 then the coefficients $\mathcal B_{\lambda,\mu}$ are given by
 \ben
 \mathcal B_{\lambda,\mu}\lb \theta_*;\sigma;\substack{\theta_t \\ \theta_0}\rb=\frac{
  Z_{\lambda,\emptyset}\lb\theta_*+\sigma\rb  Z_{\mu,\emptyset}\lb\theta_*-\sigma\rb
 \prod_{\epsilon=\pm}
  Z_{\emptyset,\lambda}\lb\epsilon\theta_0-\theta_t-\sigma\rb
  Z_{\emptyset,\mu}\lb\epsilon\theta_0-\theta_t+\sigma\rb        
              }{Z_{\lambda,\lambda}\lb \tfrac Q2\rb  
              Z_{\mu,\mu}\lb \tfrac Q2\rb Z_{\lambda,\mu}\lb \tfrac Q2+2\sigma\rb Z_{\mu,\lambda}\lb \tfrac Q2-2\sigma\rb},
  \ebn
  with $Z_{\lambda,\mu}\lb \theta\rb$ defined by (\ref{nekrofun}).
  Analogously to (\ref{classlim2f1}), the $c\to\infty$ limit of $\mathcal B\lb \theta_*;\sigma;\substack{\theta_t \\ \theta_0};t\rb$ yields  Kummer's confluent hypergeometric function,
    \ben
    \lim_{c\to\infty}\mathcal B\lb \theta_*;\sigma;\substack{\theta_t \\ \theta_0};t\rb=t^{\Delta-\Delta_0-\Delta_t}
    {}_1F_1\left(\substack{\Delta_t-\Delta_0+\Delta\\ 2\Delta};\theta_*t\right).
    \ebn
  For this reason, we are going to call $\mathcal B\lb \theta_*;\sigma;\substack{\theta_t \\ \theta_0};t\rb$ confluent CB of the 1st kind. After taking away the prefactor $t^{\Delta-\Delta_0-\Delta_t}$, this function is expected to be analytic in the entire complex $t$-plane. It inherits a part of discrete symmetries of the regular CB:
  \begin{alignat*}{3}
  \mathcal B\lb \theta_*;\sigma;\substack{\theta_t \\ \theta_0};t\rb=& \mathcal B\lb \theta_*;\sigma;\substack{\theta_t \\ -\theta_0};t\rb
  =\mathcal B\lb \theta_*;\sigma;\substack{-\theta_t \\ \theta_0};t\rb=\mathcal B\lb \theta_*;-\sigma;\substack{\theta_t \\ \theta_0};t\rb, \qquad && \text{(sign flips)} \\
    =\,& 
     t^{\frac{\theta_*^2-\lb\theta_0+\theta_t\rb^2}{2}}
     e^{\delta t}
     \mathcal B\lb \theta_*-2\delta;\sigma;\substack{
     \theta_t-\delta \\ \theta_0-\delta };t\rb, && \text{(Regge-Okamoto symmetry)}  \\   
  =&\,e^{\pm i\pi \lb \Delta-\Delta_0-\Delta_t\rb}\mathcal B\lb -\theta_*;\sigma;\substack{\theta_t \\ \theta_0};-t\rb,\quad 
  \Im t\gtrless 0,\qquad &&
  \text{(braiding transformation)} \\
        =&\, e^{\theta_* t} \mathcal B\lb -\theta_*;\sigma;\substack{\theta_0 \\ \theta_t};t\rb, \qquad && \text{(exchange of rows)}   
  \end{alignat*}
  with $\delta=\tfrac12\lb \theta_0+\theta_t+\theta_*\rb$.
  The reduced Weyl symmetry has type $ A_3$ and is generated by the flips $s_0$, $s_t$ and the Regge-Okamoto transformation $s_{\delta}$. The row exchange is not independent and can be written as $s_{\delta}s_0s_ts_{\delta}$. The table below records the transformations from $W\lb A_3\rb$.
   \begin{center}
    \begin{tabular}{c||c|c|c}
    & $\theta_0$ & $\theta_t$ & $\theta_* $ \\ \hline\hline
    $s_0$ & $-\theta_0$ & $\theta_t$ & $\theta_*$ \\ \hline
    $s_t$ & $\theta_0$ & $-\theta_t$ & $\theta_*$ \\ \hline
    $s_{\delta}$ & $\theta_0-\delta$ & $\theta_t-\delta$ & $\theta_*-2\delta$ \\ \hline\hline
    $s_r$ & $\theta_t$ & $\theta_0$ & $-\theta_*$
    \end{tabular}
    \end{center}
  
  The algebraic description of confluent CBs of the 1st kind involves the Whittaker vectors $\left\langle  \lambda_1,\lambda_2\right|$ of rank 1, that is, joint eigenstates of the Virasoro generators $L_{-1,-2}$ with eigenvalues $\lambda_{1,2}$ which annihilate all $L_{\le -3}$. There exists a canonical pairing between the Whittaker module $\mathcal W_{\lambda_1,\lambda_2}$ generated by $\left\langle  \lambda_1,\lambda_2\right|$ and the highest weight module $\mathcal M_{\Delta}$ generated by $\left|\Delta\right\rangle$. This pairing is uniquely fixed by the normalization $\left\langle  \lambda_1,\lambda_2\bigr|\Delta\right\rangle=1$. In this notation,
  \ben
  \mathcal B\lb \theta_*;\sigma;\substack{\theta_t \\ \theta_0};t\rb=\bigl\langle  \theta_*,\tfrac14\bigl|  V_{\Delta,\Delta_0}^{\Delta_t}\lb t\rb\bigr|\Delta_0\bigr\rangle\equiv\;\; 
   \begin{tikzpicture}[baseline,yshift=-0.3cm,scale=0.8]
   \draw [thick] (-1.2,-0.04) -- (0,-0.04);
   \draw [thick] (-1.2,0.04) -- (0,0.04);
   \draw [thick] (0,0) -- (2,0);
   \draw [thick] (1,0) -- (1,1); 
   \draw (-0.7,0) node[above] {\scriptsize $\lb\theta_*,\frac14\rb$};
   \draw (0.5,0) node[above] {\scriptsize $\sigma$};
   \draw (1.7,0) node[above] {\scriptsize $\theta_0$};
   \draw (1,1) node[right] {\scriptsize $\theta_t$};
   \draw [fill] (0,0) circle (0.08);
   \end{tikzpicture}\lb t\rb.
  \ebn
  The double line in the above pictorial representation  corresponds to the Whittaker module $\mathcal W_{\theta_*,\frac14}$ and the thick dot represents the pairing. The value $\lambda_2=\frac14$ carries no special meaning; it could be made arbitrary at the expense of rescaling of parameters such as $t$ and $\theta_*$.

 \subsection{Confluent conformal blocks of the 2nd kind}
 Confluent CBs of the 1st kind emerge as a limit of the regular $s$-channel CBs. They are analogous to the fundamental solutions of the Kummer's confluent hypergeometric equation associated to its regular singular point $t=0$. In its vicinity $\mathcal B\lb \theta_*;\sigma;\substack{\theta_t \\ \theta_0};t\rb$ can be efficiently computed by the convergent series expansion \eqref{confCB1series}.  It is natural to wonder if there exists a different basis/channel in the space of confluent CBs whose elements would admit asymptotic expansions  around the irregular singularity $t=\infty$, possibly inside some sectors (akin to the confluent hypergeometric function of the 2nd kind).
 
 Gaiotto and Teschner \cite{GT} proposed a procedure of taking the confluent limit of the $u$-channel regular CBs and checked that it leads to finite first terms in the asymptotic expansion at $t=\infty$. It is not difficult to extend their prescription to arbitrary order using the regular CB series \eqref{confCB1series}. The recipe boils down to consider
 \begin{align}
 \nonumber &\hat{\mathcal D}\lb\substack{\theta_t\\ \theta_*};\nu;\theta_0;t\rb\equiv \sum_{k=0}^{\infty}{\mathcal D}_k\lb\substack{\theta_t\\ \theta_*};\nu;\theta_0\rb t^{-k}:=\\
 \label{limirr}
 =&\,\lim_{\Lambda\to\infty}\lb\frac t\Lambda\rb^{
 \lb\frac{\theta_*}{2}+\nu\rb \Lambda-\frac{\theta_*^2}{4}-\Delta_t+\nu^2}\lb 1-\frac {\Lambda}t\rb^{\lb\frac{\theta_*}{2}+\nu\rb \Lambda+\frac{\theta_*^2}{4}+\Delta_t-\nu^2}
  \mathcal F\lb\substack{\frac{\Lambda+\theta_*}{2}\;\;\qquad 
               \theta_{t}\;\;\\  \frac{\Lambda}{2}+\nu \\ 
               \;\;\;\theta_0\qquad \frac{\Lambda-\theta_*}{2}};\frac{\Lambda}{t}\rb,
 \end{align}
 where $\hat{\mathcal D}\lb t\rb$ and the expression under the limit are considered as elements of $\mathbb C[[t^{-1}]]$ (formal series in $t^{-1}$). The existence of the limit is highly  nontrivial; it means that all coefficients,
 \beq\label{limirr2}
 {\mathcal D}_k\lb\substack{\theta_t\\ \theta_*};\nu;\theta_0\rb=\lim_{\Lambda\to\infty}\sum_{l=0}^k
  \sum_{\substack{\lambda,\mu\in\mathbb Y\\ |\lambda|+|\mu|=l}}
 \lb -1\rb^{k-l}{\lb\frac{\theta_*+Q}{2}+\theta_t+\nu\rb \lb\Lambda+\frac{\theta_*+Q}{2}+\theta_t-\nu\rb \choose k-l } \;
\mathcal F_{\lambda,\mu}\lb\substack{\frac{\Lambda+\theta_*}{2}\;\;\qquad 
                \theta_{t}\;\;\\  \frac{\Lambda}{2}+\nu \\ 
                \;\;\;\theta_0\qquad \frac{\Lambda-\theta_*}{2}}\rb\Lambda^l,
 \eeq
 are finite albeit the individual terms in the sum in the last limit diverge as $\Lambda^k$. To provide points of comparison for the reader, we record below the first two coefficients:
 \begin{align*}
 {\mathcal D}_1\lb\substack{\theta_t\\ \theta_*};\nu;\theta_0\rb=&\,4\nu^3-\lb 2\Delta_0+2\Delta_t+
 \theta_*^2\rb\nu +\lb \Delta_t-\Delta_0\rb \theta_*,\\
 {\mathcal D}_2\lb\substack{\theta_t\\ \theta_*};\nu;\theta_0\rb=&\,
 \tfrac12{\mathcal D}_1^2\lb\substack{\theta_t\\ \theta_*};\nu;\theta_0\rb +
 3\nu\, {\mathcal D}_1\lb\substack{\theta_t\\ \theta_*};\nu;\theta_0\rb
 -2\nu^4+\lb\Delta_t-\Delta_0\rb\theta_*\nu+\\
 +&\,\tfrac18\lb 4\Delta_0-\theta_*^2\rb \lb 4\Delta_t-\theta_*^2\rb+\frac{c-1}{12}\lb \theta_*^2-4\nu^2\rb.
 \end{align*}
 Observe that, similarly to the regular case, the central charge does not explicitly appear in $\mathcal D_1$. Assuming the existence of the limit (\ref{limirr}) to all orders in $t^{-1}$, one may deduce from the symmetries of the regular CB the  following transformations:
   \begin{alignat*}{3}
   \hat{\mathcal D}\lb\substack{\theta_t\\ \theta_*};\nu;\theta_0;t\rb=&\,
   \hat{\mathcal D}\lb\substack{\theta_t\\ \theta_*};\nu;-\theta_0;t\rb=
   \hat{\mathcal D}\lb\substack{-\theta_t\\ \theta_*};\nu;\theta_0;t\rb=\hat{\mathcal D}\lb\substack{\theta_t\\ -\theta_*};-\nu;\theta_0;-t\rb, \qquad && \text{(sign flips)} \\
     =\,& \hat{\mathcal D}\lb\substack{\theta_t-\delta\\ \theta_*-2\delta};\nu;\theta_0-\delta;t\rb
      , && \text{(Regge-Okamoto symmetry)}\\
   =&\,\hat{\mathcal D}\lb\substack{\theta_0\\ -\theta_*};\nu;\theta_t;t\rb, && \text{(exchange of momenta)}
   \end{alignat*}
   where again $\delta=\tfrac12\lb\theta_0+\theta_t+\theta_*\rb$. The last transformation can be represented as a composition $s_{\delta}s_0s_ts_{\delta}$ of reflections generating the Weyl symmetry group $W\lb A_3\rb$.
 
   An algebraic interpretation of the formal series $\hat{\mathcal D}\lb\substack{\theta_t\\ \theta_*};\nu;\theta_0;t\rb\in
   \mathbb C[[t^{-1}]]$ was suggested in \cite{Nagoya}. The idea is to extend the standard definition of the chiral vertex operator so that it intertwines two Whittaker modules. Specifically, 
   consider the operator $W_{\theta_*,\nu}^{\;\;\Delta_t}\lb t\rb:\mathcal W_{\theta_*,\frac14}\to \mathcal W_{\frac{\theta_*}{2}-\nu,\frac14}$ defined for $\nu\in\mathbb C$ by the usual commutation relations with the Virasoro generators
   \ben
   \left[L_n,W_{\theta_*,\nu}^{\;\;\Delta_t}\lb t\rb\right]=
   t^n\lb t\tfrac{\partial}{\partial t}+\lb n+1\rb\Delta_t\rb W_{\theta_*,\nu}^{\;\;\Delta_t}\lb t\rb,
   \ebn
   and the normalization condition
   \ben
   \bigl\langle  \theta_*,\tfrac14\bigl|W_{\theta_*,\nu}^{\;\;\Delta_t}\lb t\rb = t^{\frac{\theta_*^2}{2}-2\nu^2} e^{\lb \frac{\theta_*}{2}+\nu\rb t}\left[\bigl\langle  \tfrac{\theta_*}{2}-\nu,\tfrac14\bigl|+\sum_{n=1}^{\infty}t^{-n} \left\langle   w_n \right| \right],\qquad  \left\langle   w_n \right| \in \mathcal W_{\frac{\theta_*}{2}-\nu,\frac14}.
   \ebn
   The expression inside the square brackets is understood as an element of $\mathcal W_{\frac{\theta_*}{2}-\nu,\frac14}\otimes \Cb[[t^{-1}]]$. The existence and uniqueness of the operator $W_{\theta_*,\nu}^{\;\;\Delta_t}\lb t\rb$ was proved in \cite[Theorem 2.12]{Nagoya}, and was further used there to define an irregular CB
    \beq\label{algdefirrcb}
   {\mathcal D}\lb\substack{\theta_t\\ \theta_*};\nu;\theta_0;t\rb:=\bigl\langle  \theta_*,\tfrac14\bigl|  W_{\theta_*,\nu}^{\;\;\Delta_t}\lb t\rb\bigr|\Delta_0\bigr\rangle\equiv\;\;
      \begin{tikzpicture}[baseline,yshift=-0.3cm,scale=0.8]
      \draw [thick] (-1.2,-0.04) -- (2,-0.04);
      \draw [thick] (-1.2,0.04) -- (2,0.04);
      \draw [thick] (2,0) -- (3,0);
      \draw [thick] (0,0.04) -- (0,1); 
      \draw (-0.7,0) node[above] {\scriptsize $\lb\theta_*,\frac14\rb$};
      \draw (1,0) node[above] {\scriptsize $\lb\frac{\theta_*}{2}-\nu,\frac14\rb$};
      \draw (2.7,0) node[above] {\scriptsize $\theta_0$};
      \draw (0,1) node[right] {\scriptsize $\theta_t$};
      \draw [fill] (2,0) circle (0.08);
      \end{tikzpicture}\lb t\rb.
     \eeq
   We will refer to the formal series $ {\mathcal D}\lb\substack{\theta_t\\ \theta_*};\nu;\theta_0;t\rb$ as the confluent CB of the 2nd kind. It is straightforward to check that
   the first coefficients of this series coincide with those obtained with the prescription of \cite{GT}. In the present work, we show that this correspondence holds to all orders.   
   \begin{customthm}{A}\label{TheorA}
   We have 
   \ben
   {\mathcal D}\lb\substack{\theta_t\\ \theta_*};\nu;\theta_0;t\rb
   =t^{\frac{\theta_*^2}{2}-2\nu^2} e^{\lb \frac{\theta_*}{2}+\nu\rb t}\sum_{k=0}^{\infty}{\mathcal D}_k\lb\substack{\theta_t\\ \theta_*};\nu;\theta_0\rb t^{-k}.
   \ebn
   \end{customthm}
   The benefits of this identification are two-fold. First, it shows the consistency of the limit (\ref{limirr}) and ensures that all the coefficients  ${\mathcal D}_k\lb\substack{\theta_t\\ \theta_*};\nu;\theta_0\rb$ are indeed finite. On the other hand, it allows to replace their rather cumbersome computation based on the algebraic definition (\ref{algdefirrcb}) by employing (\ref{limirr2}) together with the explicit coefficients (\ref{rCBcoef}) of the regular CB expansion coming from the AGT correspondence. The latter approach is much easier to implement in computer algebra systems.

 \subsection{Painlevé V asymptotics and connection problems\label{subsecintropv}}
  The regular CBs with Virasoro central charge $c=1$ are intimately related to the Painlevé VI transcendents \cite{GIL12,ILTe,BS}. This relation will be explained in detail in the main body of the paper. The irregular CBs of two kinds discussed above are expected to be connected in a similar manner to Painlevé V  \cite{GIL13,Nagoya,Nagoya2}. It turns out that the careful study of the confluence of CBs combined with the known Painlevé VI results enable one to solve a connection problem for Painlevé~V (PV) which we now briefly describe.  
  
  The PV tau function $\tau\lb t\rb$ is most frequently defined by its logarithmic derivative $t\tfrac{d}{dt}\ln\tau= H+\frac{\theta_*\lb t+\theta_*\rb}{2}$, where $H\lb t\rb$ denotes the non-autonomous PV Hamiltonian and satisfies 
  \beq\label{sigpv}
  \lb t\ddot H\rb^2=\lb H-t\dot H+2\dot H^2\rb^2-\tfrac14\lb\lb2 \dot H-\theta_*\rb^2-4\theta_0^2\rb \lb \lb2 \dot H+\theta_*\rb^2-4\theta_t^2\rb,
  \eeq
  The dots here and below denote derivatives with respect to $t$.
  Painlevé V describes isomonodromic deformations of rank 2 linear systems with two Fuchsian and one irregular singularity of Poincaré rank 1. The generalized monodromy data encode simultaneously the parameters $\theta_{0,t,*}$  and the initial conditions.
  
  The Painlevé property means that $H\lb t\rb$ meromorphically continues to the universal cover $\widetilde{\mathbb C\backslash\left\{0\right\}}$. Likewise, the tau function $\tau\lb t\rb$ is holomorphic on $\widetilde{\mathbb C\backslash\left\{0\right\}}$. To make $\tau\lb t\rb$ single-valued on the base, it is convenient to introduce the branch cut along $\mathbb R_{\leq 0}$ connecting the branch points $t=0,\infty$.\vspace{0.2cm}\\
  \textbf{1.4.1. Asymptotics as $t\to 0$}.
 The short-distance asymptotic behavior of $\tau\lb t\rb$  has been obtained  in \cite[Theorem 3.1]{Jimbo}. This result was upgraded to a complete expansion at $t=0$ by Conjecture 3 of \cite{GIL13}, the proof of which will be given in the present work.
 \begin{customthm}{B}\label{theoB} Let $\sigma,\eta\in\Cb$ satisfy the condition
 $2\sigma\notin\mathbb Z$.
  The Painlevé V tau function $\tau\lb t\rb$ corresponding to generic monodromy admits the following  representation in the neighborhood of $t=0$:
 \begin{subequations}
 \beq
 \label{fourier0}
 \tau\lb t\rb =\mathcal N_0\sum_{n\in\mathbb Z} e^{2\pi in\eta}
 \mathcal C_0\lb \theta_*;\sigma+n;\substack{\theta_t \\ \theta_0}\rb \mathcal B\lb \theta_*;\sigma+n;\substack{\theta_t \\ \theta_0};t\rb\Bigl|_{c=1},
 \eeq
 where  $\sigma,\eta$ correspond to the initial conditions, $\mathcal B\lb \theta_*;\sigma;\substack{\theta_t \\ \theta_0};t\rb$ is given by the combinatorial series \eqref{confCB1series} (with $c=1$) and the structure constants $\mathcal C_0\lb \theta_*;\sigma;\substack{\theta_t \\ \theta_0}\rb$ are expressed in terms of the Barnes $G$-function as
 \beq\label{strfourier0}
 \mathcal C_0\lb \theta_*;\sigma;\substack{\theta_t \\ \theta_0}\rb=\prod_{\epsilon=\pm}\frac{G\lb 1+\theta_*+\epsilon\sigma\rb G\lb 1+\theta_0+\theta_t+\epsilon\sigma\rb G\lb 1-\theta_0+\theta_t+\epsilon\sigma\rb}{G\lb 1+2\epsilon\sigma\rb}.
 \eeq
 \end{subequations}
 \end{customthm}
 
 The asymptotic behavior of generic Painlevé V functions as $t\to\infty$ is considerably more intricate. Thanks to the symmetry $\lb H,t,\theta_*\rb\mapsto \lb H,-t,-\theta_*\rb$ of the equation (\ref{sigpv}) it may be assumed that $0\leq \arg t<\pi$. On the rays $\arg t=0,\frac{\pi}{2}$ the long-distance asymptotics is expressed in terms of powers of $t$ and exponentials. To authors' knowledge, the $t\to\infty$ asymptotics inside the sectors $0<\arg t<\frac{\pi}{2}$ and $\frac{\pi}{2}<\arg t<\pi$ has not been studied in the literature. Its counterparts for Painlevé I and II involve modulated elliptic functions \cite{JK,KK,FIKN}. Recent results of \cite{JL} suggest that this might also be the case for Painlevé V.\vspace{0.2cm}\\
 \textbf{1.4.2. Asymptotics as $t\to i\infty$}. 
 The study of the asymptotics of  $\tau\lb t\rb$ as $t\to i\infty$ beyond the leading order reveals that the corresponding asymptotic expansion can be written in a form similar to
 (\ref{fourier0}), but with $c=1$ confluent CBs of the 2nd kind instead of the 1st. Specifically, it was conjectured in \cite{Nagoya} that the generic 2-parameter family of PV tau functions is characterized by the expansion
 \begin{subequations}
 \beq \label{fourier8}
 \tau\lb t\to i\infty\rb\simeq \mathcal N_{i\infty}\sum_{n\in\mathbb Z} e^{2\pi  in\rho}
 \mathcal C_{i\infty}\lb\substack{\theta_t\\ \theta_*};\nu+n;\theta_0\rb {\mathcal D}\lb\substack{\theta_t\\ \theta_*};\nu+n;\theta_0;t\rb\Bigl|_{c=1},
 \eeq
 where ${\mathcal D}\lb\substack{\theta_t\\ \theta_*};\nu;\theta_0;t\rb$ is given by \eqref{algdefirrcb} and can be efficiently computed using Theorem~\ref{TheorA} and the limit (\ref{limirr2}). The structure constants are identified as
 \beq \label{strfourier8}
 \mathcal C_{i\infty}\lb\substack{\theta_t\\ \theta_*};\nu;\theta_0\rb
 =\lb2\pi\rb^{-2\nu}\prod_{\epsilon=\pm}G\lb 1+\nu+\epsilon\theta_0-\tfrac{\theta_*}{2}\rb  
 G\lb 1+\nu+\epsilon\theta_t+\tfrac{\theta_*}{2}\rb,
 \eeq
 and the initial conditions are encoded in the parameters $\nu,\rho\in\mathbb C$. In contrast to the series \eqref{fourier0}, expected to converge for $t\in\Cb\backslash \mathbb R_{\leq0}$, the  expansion \eqref{fourier8} is only asymptotic. Another possibility is to interpret \eqref{fourier8} as an exact equality defining the $c=1$ confluent CBs of the 2nd kind non-perturbatively. \vspace{0.2cm}\\
 \end{subequations}
 \textbf{1.4.3. Asymptotics as $t\to +\infty$}. It was observed in \cite{BLMST} that the ``Fourier transform'' structure of  \eqref{fourier0} and \eqref{fourier8} also shows up in the PV long-distance asymptotic series on the real axis, as well as for all the other Painlevé tau functions on the appropriate canonical rays\footnote{This surprising periodicity phenomenon was first observed (in the irregular case) for Painlevé III equation of type $D_8$ in \cite{ILT14}, and used there to obtain the connection constant for the corresponding tau function.}. The proposal for PV is that
 \begin{subequations}
 \begin{align}\label{fourier8r}
 &\tau\lb t\to +\infty\rb=  \mathcal N_{+\infty}\sum_{n\in\mathbb Z}e^{2\pi in\xi} \mathcal C_{+\infty}\lb\omega+n\rb\mathcal G\lb\substack{\theta_t\\ \theta_*};\omega+n;\theta_0;t\rb,\\
 & \mathcal C_{+\infty}\lb\omega\rb=   2^{-4\theta_0^2-4\theta_t^2-2\theta_*^2-\frac{{\omega}^2}{2}}\lb2\pi\rb^{-\frac{\omega}{2}}e^{-\frac{\pi i {\omega}^2}{4}}G\lb 1+\omega\rb,
 \end{align}
 where $\omega,\xi\in\mathbb C$ parameterize the initial conditions and $\mathcal G\lb \substack{\theta_t\\ \theta_*};\omega;\theta_0; t\rb$ has the following asymptotic expansion as $t\to +\infty$:
 \beq
 \begin{aligned}
 \mathcal G\lb\substack{\theta_t\\ \theta_*};\omega;\theta_0;t\rb
 \simeq  t^{\theta_0^2+\theta_t^2+\theta_*^2-\frac{{\omega}^2}{2}-\frac14} e^{\frac{t^2}{32}+\frac{\lb i\omega+\theta_*\rb t}{2}}\left[1+\sum_{k=1}^{\infty}\mathcal G_k\lb\substack{\theta_t\\ \theta_*};\omega;\theta_0\rb t^{-k}\right].
 \end{aligned}
 \eeq
 The coefficients $\mathcal G_k$ can be recursively calculated  up to any finite order from the PV equation \eqref{sigpv}. As of now, we do not have an explicit general formula for them nor an independent algebraic definition of $ \mathcal G\lb t\rb$, i.e. the analogs of \eqref{limirr2} and \eqref{algdefirrcb} are yet to be found. \vspace{0.2cm}
 \end{subequations}
 
 The classical connection problem for Painlevé V is to explicitly relate the different pairs of  parameters such as $\lb\sigma,\eta\rb$, $\lb \nu,\rho\rb$ and $\lb \omega,\xi\rb$ characterizing the behavior of
 PV functions as $t\to 0$, $t\to i\infty$ and $t\to +\infty$  (as well as parameters of the asymptotics inside the sectors
 $\arg t\in\lb 0,\frac{\pi}{2}\rb$, $\lb \frac{\pi}{2},\pi\rb$ once it becomes available). A possible approach to Painlevé connection problems consists in expressing all of the asymptotic parameters in terms of monodromy data. It was first implemented by Jimbo \cite{Jimbo} for Painlevé VI. In the PV case, he also related to monodromy the short-distance asymptotic parameters $\lb \sigma,\eta\rb$. The corresponding task for parameters $\lb\omega,\xi\rb$ at $t\to+\infty$ was accomplished by Andreev and Kitaev in \cite{AK}, which allowed them to solve the PV connection problem between $0$ and $+\infty$. We are going to use the confluence of CBs to provide a conjectural solution of the connection problem between $0$ and $i\infty$. These results are summarized below for reader's convenience\footnote{The connection problems between $0$, $+\infty$ and $i\infty$ were also studied in \cite{MT1,MT2,MT3} and more recently in e.g. \cite{AK2,DKV,ZZ,LZZ} for special classes of PV~solutions with constrained parameters $\theta_{0,t,*}$ and non-generic initial conditions. The present paper is almost exclusively concerned with PV functions corresponding to monodromy in general position.}.
 \vspace{0.2cm}\\
 \textbf{1.4.4. Connection formulae between $0$ and $+\infty$}. 
 Define the following combinations of Painlevé V parameters:
   \beq
   \label{Apm}
   \begin{aligned}
   A_{\pm}:=&\,2e^{\mp\pi i \theta_*}\cos2\pi\theta_0+2e^{\pm\pi i \theta_*}\cos2\pi\theta_t.
   \end{aligned}
   \eeq
 It it straigtforward to check that they are invariant under the $W\lb A_3\rb$ action.
 Let $\lb\sigma,\eta\rb$ be the parameters of the short-distance PV tau function expansion (\ref{fourier0}) satisfying the conditions of Theorem~\ref{theoB}. Introduce auxiliary quantities $X_{\pm}$ by
      \begin{align} \label{zetap}
      2X_{\pm}\sin^22\pi\sigma=&\, A_{\pm}-A_{\mp}\cos2\pi\sigma\mp 2i\sum_{\epsilon=\pm}
      e^{2i\pi\epsilon\eta\mp i\pi\epsilon\sigma}
      \lb\cos2\pi\lb\theta_t-\epsilon\sigma\rb-\cos2\pi\theta_0\rb
      \sin\pi\lb \theta_*-\epsilon\sigma\rb.
      \end{align}
   Under the genericity assumptions $X_+X_-\ne 1$,
   $X_+\ne 0$, the parameters $\lb\omega,\xi\rb$ of the long-distance expansion  \eqref{fourier8r} on the real axis are given by
   \beq\label{nup}
   e^{2\pi i\omega}=1-X_+X_-,\qquad e^{2\pi i\xi}=X_+.
   \eeq
   The identification \eqref{zetap}--\eqref{nup} is essentially a compact paraphrase of the results of \cite[Sections 3--4]{AK} corresponding to generic strata in the space of monodromy data.
      \vspace{0.2cm}\\
 \textbf{1.4.5. Connection formulae between $0$ and $i\infty$}. 
 Assuming the genericity conditions $X_+X_-\ne1$, $X_-\ne 0$, we put forward the following claim.
 \begin{customconj}{C}
 The asymptotic parameters $\lb\sigma,\eta\rb$ and $\lb\nu,\rho\rb$ of the short-distance expansion (\ref{fourier0}) and long-distance expansion (\ref{fourier8}) on the imaginary axis are related by
 \beq\label{nudefi}
 e^{2\pi i \nu}=X_-,\qquad e^{2\pi i \rho}=1-X_+X_-,
 \eeq
 where $X_{\pm}$ are defined by the same formula (\ref{zetap}).
 \end{customconj}
 \noindent\textbf{1.4.6. Connection constants for $\tau\lb t\rb$ between $0$,  $i\infty$ and $+\infty$}. The amplitudes $\mathcal N_0$, $\mathcal N_{i\infty}$, $\mathcal N_{+\infty}$ in the asymptotic series (\ref{fourier0}), (\ref{fourier8}), (\ref{fourier8r})  are intrinsically undetermined. This is due to the normalization ambiguity of the PV tau function defined by its logarithmic derivative. However, the ratios
 \beq
   \Upsilon_{0\to i\infty}:=\frac{\mathcal N_{i\infty}}{\mathcal N_0},\qquad
 \Upsilon_{i\infty\to +\infty}:=\frac{\mathcal N_{+\infty}}{\mathcal N_{i\infty}}, 
 \eeq
 describing relative normalizations of three expansions,
 are completely fixed by the Painlevé V equation and appropriate initial conditions/monodromy data. The evaluation of these connection constants constitutes one of the most subtle asymptotic problems in the Painlevé theory, relevant for many applications (see e.g. recent papers \cite{ILP,BIP,ILT14,IP1,IP2,LR,Bot,DKV} and references to earlier works therein).
  \begin{customconj}{D}
  Denote $\hat G\lb z\rb:=\ds\frac{G\lb 1+z\rb}{G\lb 1-z\rb}$ and introduce a parameter $\lambda$ by
  \ben
  e^{2\pi i \lambda }:=
  \frac{e^{i\pi \lb 2\theta_0-\theta_*\rb}+e^{i\pi\lb2\theta_t+\theta_*\rb}-X_+-e^{-2\pi i \sigma}X_-}{e^{2i\pi\lb\theta_0+\theta_t+\sigma\rb}-1}.
  \ebn
  Then the relative normalizations are given by
  \begin{subequations}\label{Upss}
  \beq   \label{Ups0i8}
  \begin{aligned}
  \Upsilon_{0\to i\infty}\;\;=&\,\frac{\lb2\pi\rb^{2\nu} e^{i\pi\nu\lb\nu-\sigma\rb+
  i\pi\lb\sigma+\theta_0-\frac{\theta_*}{2}\rb\lb\sigma+\theta_t+\frac{\theta_*}{2}\rb}
  \hat{G}\lb \sigma-\theta_0+\theta_t\rb
    \hat{G}\lb\sigma+\theta_*\rb}{\hat{G}\lb \sigma-\theta_0-\theta_t\rb
      \hat{G}\lb\nu-\theta_0-\frac{\theta_*}{2}\rb
       \hat{G}\lb\nu-\theta_t+\frac{\theta_*}{2}\rb}\times\\
  \times &\,\frac{
  e^{i\pi\lb\theta_0+\theta_t+\sigma\rb\lambda}
   \hat{G}\lb \lambda-\nu+\sigma\rb
   \hat{G}\lb \lambda+\theta_0+\frac{\theta_*}{2}\rb
  \hat{G}\lb \lambda+\theta_t-\frac{\theta_*}{2}\rb
  }{
  \hat{G}\lb\lambda-\nu+\theta_0+\theta_t\rb
  \hat{G}\lb \lambda+\sigma+\theta_0-\frac{\theta_*}{2}\rb\hat{G}\lb \lambda+\sigma+\theta_t+\frac{\theta_*}{2}\rb
  },\end{aligned}
  \eeq
  \beq
  \label{Upsi88}
  \Upsilon_{i\infty\to +\infty}= G^2\lb\tfrac12\rb e^{-\frac{i\pi}{24}}
  \lb 2\pi\rb^{\omega+\frac12}e^{\frac{\pi i \omega^2}{2}-2\pi i\omega\nu}\hat{G}\lb -\omega\rb.
  \eeq
  \end{subequations}
  \end{customconj}
  
  Even though both evaluations are conjectural, they are arrived at in different ways. The formula for  $\Upsilon_{0\to i\infty}$ is obtained as a limit of the Painlevé VI connection constant \cite{ILT13,ILP} under a prescription analogous to the one used above to define confluent CBs of the 2nd kind. Lacking a similar CFT interpretation of the real axis expansions, we found (\ref{Upsi88})  by employing instead the recurrence relations for $\Upsilon_{i\infty\to +\infty}$ with respect to monodromy parameters $\nu$, $\omega$. An additional numerical guesswork was necessary to fix the part of   $\Upsilon_{i\infty\to +\infty}$ independent of $\nu$ and $\omega$. We hope to provide a rigorous proof of Conjectures~C and~D in a future work.

  \section{Confluent conformal blocks}
   
   Let $\mathcal M_\Delta$ and $\mathcal M_\Delta^*$ be a Verma module of the Virasoro algebra 
   with the highest weight $\Delta$ and its dual. We normalize the canonical bilinear pairing $\left\langle |\right\rangle: 
   \mathcal M^*_\Delta \times \mathcal M_\Delta\to \mathbb{C}$ as $\left\langle \Delta|\Delta\right\rangle=1$. 
   Denote by $V_{\Delta_1,\Delta_3}^{\Delta_2}\lb z\rb:\mathcal M^*_{\Delta_1}
      \to \mathcal M^*_{\Delta_3}$ the primary chiral vertex operator of weight $\Delta_2$ intertwining two Verma modules $\mathcal M^*_{\Delta_1}$ and $\mathcal M^*_{\Delta_3}$. It satisfies the following commutation  relations with the Virasoro generators:
   \begin{equation}\label{eq_chiral_VO1}
   \left[ L_n, V_{\Delta_1, \Delta_3}^{\Delta_2}\lb z\rb
   \right]=z^n\left( z \tfrac{\partial}{\partial z}+
   \lb n+1\rb\Delta_2\right)
   V_{\Delta_1, \Delta_3}^{\Delta_2}\lb z\rb,
   \end{equation}
  and is normalized as 
  \begin{equation}\label{eq_chiral_VO2}
  \left\langle \Delta_1\right|
  V_{\Delta_1, \Delta_3}^{\Delta_2}\lb z\rb
  =z^{\Delta_1-\Delta_2-\Delta_3}
  \left[ \left\langle \Delta_3\right|+\sum_{n=1}^\infty z^{-n}\left\langle v_n\right|\right],\qquad 
  \left\langle v_n\right|\in \mathcal M^*_{\Delta_3}. 
  \end{equation} 
   From the definitions  \eqref{eq_chiral_VO1} and \eqref{eq_chiral_VO2}, we have 
   \begin{subequations}
   \begin{align}
   \left\langle v_n\right|L_{0}=&\left(\Delta_3+n\right)
   \left\langle v_n\right|, 
   \\
   \left\langle v_n\right|L_{-1}=&\left( \Delta_3+\Delta_2-\Delta_1+n-1 \right)\left\langle v_{n-1}\right|,
   \label{eq_v_L-1}
   \\
    \left\langle v_n\right|L_{-2}=&\left(\Delta_3+2\Delta_2-\Delta_1+n-2  \right)\left\langle v_{n-2}\right|. 
    \label{eq_v_L-2}
   \end{align}
   \end{subequations}
  These three relations determine all $\left\langle v_n\right|$ uniquely if $\mathcal M^*_{\Delta_1}$ 
  is irreducible. 
   The structure of relations \eqref{eq_v_L-1} and \eqref{eq_v_L-2} implies that we can take a limit of 
  $z^{-n}\left\langle v_n\right|$ as $z=\Lambda\to \infty$
  if
  \begin{equation*}
  \Delta_2-\Delta_1=\sum_{m=0}^\infty a_m\Lambda^{1-m},\qquad
  2\Delta_2-\Delta_1=\sum_{m=0}^\infty b_m\Lambda^{2-m} 
  \end{equation*}
   with some $a_m,b_m\in\Cb$. 
   
   For example, a confluence limit of \eqref{eq_chiral_VO2} as $z\to \infty$ is realized by 
   \begin{equation}\label{eq_con_limit_IVO}
   \Delta_1=\frac{c-1}{24}+\frac{\left(\Lambda-\theta_*\right)^2}{4},\qquad 
   \Delta_2=\frac{c-1}{24}+\frac{\left(\Lambda+\theta_*\right)^2}{4},\qquad
   z=\Lambda,\qquad \Lambda\to \infty, 
   \end{equation}
   which is equivalent to \eqref{conflimit}. 
   Setting
   \begin{equation*}
   \left\langle I\right|=\lim_{\Lambda\to \infty} \Lambda^{-\Delta_1+\Delta_2+\Delta_3}
   \left\langle \Delta_1\right|
  V_{\Delta_1, \Delta_3}^{\Delta_2}\lb \Lambda\rb, 
   \end{equation*} 
   we deduce from the definition \eqref{eq_con_limit_IVO} that 
   \begin{equation*}
   \left\langle I \right| L_{-1}=\theta_*\left\langle I \right|,\qquad \left\langle I\right|L_{-2}=\tfrac{1}{4}\left\langle I \right|,
   \qquad \left\langle I\right|L_{n<-2}=0,
   \qquad \left\langle I|\Delta_3\right\rangle=1. 
   \end{equation*}
   Therefore, $\left\langle I\right|$ may be viewed as a Whittaker vector $\left\langle \theta_*, \frac{1}{4}\right|$ 
   of rank $1$. 
   Since 
  \begin{equation*}
  z^{\Delta_0+\Delta_t+\Delta_1-\Delta_\infty}\left\langle \Delta_{\infty}\right|V_{\Delta_{\infty},\Delta}^{\Delta_1}\lb z\rb V_{\Delta,\Delta_0}^{\Delta_t}\lb t_1\rb\left|\Delta_0\right\rangle
  =\left\langle \Delta_{\infty}\right|V_{\Delta_{\infty},\Delta}^{\Delta_1}\lb 1\rb V_{\Delta,\Delta_0}^{\Delta_t}\lb t\rb\left|\Delta_0\right\rangle,
  \end{equation*} 
  with $t=t_1/z$,  the limit \eqref{eq_con_limit_IVO} corresponds to the confluent limit of 
   $ \mathcal F\lb\substack{\theta_{1}\;\quad 
    \theta_{t}\\ \sigma \\ \theta_{\infty}\quad \theta_0};t\rb$
  described by \eqref{conflimit}.

   On the other hand, taking the confluence limit of the $u$-channel regular CB corresponds to 
   considering a vector
   $
   \left\langle \Delta_1 \right| V_{\Delta_1, \Delta_3}^{\Delta_2}\lb t_1\rb
   V_{\Delta_3, \Delta_5}^{\Delta_4}\lb z\rb
   $ 
   as $z\to \infty$. The first term of its expansion in $t_1$ admits the limit under the prescription
   \eqref{eq_con_limit_IVO}. The second term, however, diverges 
   even if we let $\Delta_2$, $\Delta_3$ go to $\infty$. 
  One of the reasons is that  for the regular CB $
   \left\langle \Delta_1 \right| V_{\Delta_1, \Delta_3}^{\Delta_2}\lb t_1\rb
   V_{\Delta_3, \Delta_5}^{\Delta_4}\lb z\rb\left|\Delta_5\right\rangle
   $  we assume $\left|t_1\right|>|z|$ and letting $z\to \infty$ breaks this assumption. 
   Instead, Gaiotto and Teschner \cite{GT}
   proposed the following rearranged expansion:
   \begin{equation}\label{eq_rearranged_expansion}
   \left\langle \Delta_1 \right| V_{\Delta_1, \Delta_3}^{\Delta_2}\lb t_1\rb
   V_{\Delta_3, \Delta_5}^{\Delta_4}\lb z\rb
   =t_1^{\Delta_1-\Delta_2-\Delta_3}z^{\Delta_3-\Delta_4-\Delta_5}
    \lb 1-\frac{z}{t_1}\rb^A \sum_{k=0}^\infty 
   t_1^{-k} \left\langle R_k\right|,\qquad \left\langle R_k\right|\in \mathcal M^*_{\Delta_5}
   \end{equation}
   for a suitably chosen constant $A$. 
  Here the prefactor is interpreted as an element of $\mathbb C[[z/t_1]]$, i.e.
  \begin{equation}
  \lb 1-\frac{z}{t_1}\rb^A\equiv\sum_{i=0}^\infty \frac{A\lb A-1\rb\cdots \lb A-n+1\rb}{n!}
  \lb-\frac{z}{t_1}\rb^n. 
  \end{equation}
   The above rearranged expansion is inspired by how 
   we can take limits of local solutions of the Gauss hypergeometric equation. 
   
  \begin{rmk} 
  The Gauss hypergeometric equation
  \begin{equation}
  x(1-x)\frac{d^2y}{dx^2}+\left[ \gamma-(\alpha+\beta+1)x\right]\frac{dy}{dx}-\alpha \beta y=0
  \end{equation}
  has linearly independent local solutions
  \begin{subequations}
  \begin{equation}
  {}_2F_1\left(\substack{\alpha,\beta\\ \gamma};x\right),\qquad
  x^{1-\gamma}{}_2F_1\left(\substack{\alpha-\gamma+1,\beta-\gamma+1\\2-\gamma};x\right)
  \end{equation}
  at $x=0$ and 
  \begin{equation}
  x^{-\alpha}{}_2F_1\left(\substack{\alpha,\alpha-\gamma+1\\ \alpha-\beta+1};x^{-1}\right),\qquad
  x^{-\beta}{}_2F_1\left(\substack{\beta-\gamma+1,\beta\\\beta-\alpha+1};x^{-1}\right)
  \end{equation}
  \end{subequations}
  at $x=\infty$. The series parts of the first three local solutions naturally  
  admit confluence limits by replacing $x$ with $x_1$ as $
  x=\beta^{-1}{x_1}$
  and letting $\beta\to \infty$.  Meanwhile, the coefficients of $x_1^{-n}$ ($n>0$) in the series part 
  ${}_2F_1\left(\substack{\beta-\gamma+1,\beta\\\beta-\alpha+1};\beta x_1^{-1}\right)$ 
  in the last local solution diverge. 
  From the list of Kummer's 24 solutions for the Gauss hypergeometric equation we have
  \begin{equation}
  x^{-\beta}{}_2F_1\left(\substack{\beta-\gamma+1,\beta\\\beta-\alpha+1};x^{-1}\right)
  =\lb -x\rb^{\alpha-\gamma}\lb 1-x\rb^{\gamma-\alpha-\beta}
  {}_2F_1\left(\substack{1-\alpha,\gamma-\alpha\\ \beta+1-\alpha};x^{-1}\right). 
  \end{equation}
  Then, $\lb 1- \beta^{-1}x_1\rb^{\gamma-\alpha-\beta}$ and the coefficients of 
  $x_1^{-n}$ ($n>0$) in the series part 
  ${}_2F_1\left(\substack{1-\alpha,\gamma-\alpha\\ \beta+1-\alpha};\beta x_1^{-1}\right)$ 
  admit limits as $\beta\to\infty$. In this way, we may obtain Kummer's confluent hypergeometric 
  functions of the 2nd kind as limits of local solutions of the Gauss hypergeometric equation   at $\infty$. 
   \end{rmk}
   
   \begin{prop}\label{prop_limit_VO}
  A composition of two vertex operators intertwining Verma modules 
   $V_{\Delta_1, \Delta_3}^{\Delta_2}\lb t\rb
   V_{\Delta_3, \Delta_5}^{\Delta_4}\lb z\rb: \mathcal M_{\Delta_1}^*\to 
   \mathcal M_{\Delta_5}^*$
  degenerates to a vertex operator $W_{\theta_*,\nu}^{\;\;\Delta_t}\lb t\rb:\mathcal W_{\theta_*,\frac14}\to \mathcal W_{\frac{\theta_*}{2}-\nu,\frac14}$ 
  intertwining Whittaker modules 
  by the limit 
  \begin{subequations}\label{whlim}
  \begin{alignat}{3}
  &\Delta_1=A+\Delta_3+\Delta_t+\frac{\theta_*^2}{2}-2\nu^2, \qquad 
  &&\Delta_2=\Delta_t,
  \label{eq_limit-rankone1}
  \\
  &
  \Delta_3=\frac{c-1}{24}+\frac{\left(\Lambda-\theta_*/2+\nu\right)^2}{4},\qquad 
   &&\Delta_4=\frac{c-1}{24}+\frac{\left(\Lambda+\theta_*/2-\nu\right)^2}{4},\qquad
  \label{eq_limit-rankone2}
   \\
   &A=-\lb \frac{\theta_*}{2}+\nu\rb\Lambda,\qquad 
    z=\Lambda,\qquad &&\Lambda\to \infty.\label{eq_limit-rankone3}
  \end{alignat}
  \end{subequations}
  \end{prop}
   \pf
   We are going to show that the vectors $\left\langle R_k\right|$ for $k\ge 0$ given by \eqref{eq_rearranged_expansion} 
   converge in the limit \eqref{whlim}. 
   
   By definition, $\left\langle R_0\right|$ is the series part of $ \left\langle \Delta_3 \right| 
   V_{\Delta_3, \Delta_5}^{\Delta_4}\lb z\rb$, namely, 
   $\left\langle R_0\right|=z^{-\Delta_3+\Delta_4+\Delta_5}
   \left\langle \Delta_3 \right| 
   V_{\Delta_3, \Delta_5}^{\Delta_4}\lb z\rb$. 
   Then $\left\langle R_0\right|$ goes to the Whittaker vector $\bigl\langle \tfrac{\theta_*}{2}-\nu, \frac{1}{4}\bigr|$ of rank $1$    as $\Lambda\to \infty$. 
     For a partition $\lambda=\lb \lambda_1,\ldots,\lambda_i\rb \in\mathbb Y$,  we set $L_\lambda=L_{\lambda_1-1}\cdots L_{\lambda_i-1}$. 
   By the commutation relations \eqref{eq_chiral_VO1},
   the vectors $\left\langle R_k\right|$ with $k\ge 0$ satisfy the following relations:
   \begin{align}
   \left\langle R_k\right|\lb L_n+z^n\left( \Delta_3+n\Delta_4-k-L_0\right)\rb=&A\sum_{m=1}^{-n-1}z^{n+m}
   \left\langle R_{k-m}\right|
   -\left(-A-k-n+\Delta_1+n\Delta_2-\Delta_3 \right)
   \left\langle R_{k+n}\right|, \label{eq_RL_relations}
   \end{align}
   for any $n<0$. 
   Also, we can express the vectors $\left\langle R_k\right|$  as 
   linear combinations $\sum_{\lambda\in\mathbb{Y}}r_\lambda^{(k)} \left\langle R_0\right| L_\lambda$, 
   where the sum is finite and $r_\lambda^{(k)}$ is a rational function of 
   $\Delta_{1,\ldots,5}$,  $A$, $z$ and $c$. 
   
  It is easy to see that the coefficients of $\left\langle R_{k-m}\right|$ (with $m=0,1,\ldots,-n$) in \eqref{eq_RL_relations} 
  converge in the limit  \eqref{whlim}. 
  Now suppose that $\langle R_j|$ with $j<k$ converge. From \eqref{eq_RL_relations} 
  it is not obvious whether  
   $\left\langle R_k\right|$ 
   admit limits, since the action of $ L_n-z^n\left(k+L_0\right)$ on $\left\langle R_k\right|$ may kill 
   the divergent part of the latter. 
   
   The limits of the relations \eqref{eq_RL_relations} are given by (recall that $n<0$)
   \begin{align}
   \left\langle w_k\right| L_n=&-\lb \tfrac{\theta_*}{2}+\nu\rb\langle w_{k+n+1}|
   -\lb \lb n+1\rb\Delta_t+\tfrac{\theta_*^2}{2}-2\nu^2-k-n\rb
   \left\langle w_{k+n}\right|
   +\lb \theta_*\delta_{n,-1}+\tfrac{1}{4}\delta_{n,-2}\rb
   \left\langle w_k\right|,   
  \label{eq_wL_relations} 
   \end{align}  
  and are the same as the recurrence relations for $\left\langle w_k\right|$ in the decomposition of the Whittaker vector 
  $\bigl\langle \theta_*, \frac{1}{4}\bigr| W_{\theta_*,\nu}^{\;\;\Delta_t}\lb t\rb$. 
  Indeed, the vectors $\left\langle w_k\right|$ are also written as 
  $\left\langle w_k\right|=\sum_{|\lambda|\le k}c_\lambda^{(k)} \bigl\langle \frac{\theta_*}{2}-\nu, 
  \frac{1}{4}\bigr|L_\lambda$. 
   In \cite[Theorem~2.12]{Nagoya} it was shown that 
   the relations \eqref{eq_wL_relations} give the recursive relations for $c_\lambda^{(k)}$ which uniquely determine these coefficients  as polynomials in $\theta_*$, $\nu$, $\Delta_t$, and the central charge $c$.  
   Thus  
  only $z^n\lb k+L_0\rb$ may produce a divergent part $O\lb\Lambda^{-n-1}\rb$ in $\left\langle R_k\right|$.  
   
   Therefore, $\left\langle R_k\right|$ converges in the limit 
   \eqref{whlim} because $n<0$.  
   \epf
  \begin{rmk}
   From the above proof,  we see that the uniqueness of $\left\langle w_k\right|$ in the decomposition of  
   $\bigl\langle \theta_*, \frac{1}{4}\bigr| W_{\theta_*,\nu}^{\;\;\Delta_t}\lb t\rb$ implies the existence of the limit of $\left\langle R_k\right|$. Then,  $\left\langle w_k\right|$ exist as limits of $\left\langle R_k\right|$. Therefore, 
   we get another proof of existence of the irregular vertex operator 
   $\bigl\langle \theta_*, \frac{1}{4}\bigr| W_{\theta_*,\nu}^{\;\;\Delta_t}\lb t\rb$, different from the proof 
   of \cite[Theorem 2.12]{Nagoya}. 
   \end{rmk}
   
   Theorem A is now obtained from Proposition \ref{prop_limit_VO} by projecting the composition  $V_{\Delta_1, \Delta_3}^{\Delta_2}\lb t\rb
      V_{\Delta_3, \Delta_5}^{\Delta_4}\lb z\rb$ on the highest weight states and taking the limit \eqref{whlim}.

 \section{Painlevé V}
 \subsection{Preliminaries}
 \textbf{3.1.1. Lax pair}.  The Lax pair for Painlevé V is defined by
 \begin{subequations}
  \begin{numcases}{}\label{linsys}\partial_{z}\Phi=\Phi A\lb z,t\rb=
  \Phi\lb\ds\frac{t\sigma_z}{2}+\frac{ A_0}{z}+\frac{ A_1}{z-1}
  \rb,\vspace{0.1cm}\\
  \partial_{t}\Phi\,=\Phi B\lb z,t\rb=\Phi\lb\frac{\ds z\sigma_z}{2}+ B_0\rb,
  \end{numcases}
  \end{subequations}
  where $A_0,A_1,B_0\in\mathfrak{sl}_2\lb\mathbb C\rb$ are parameterized by three scalar quantities $x$, $y$, $u$ (to be later considered as functions of~$t$) and three constant parameters $\theta_0$, $\theta_t$, $\theta_{*}$:
  \begin{align*}
  A_0=&\,\lb\begin{array}{cc} x & \frac{x-\theta_0}{u} \\ 
 -u\lb x+\theta_0\rb  & -x\end{array}\rb,\qquad
  A_1=\lb\begin{array}{cc} \theta_*-x & 
 - \frac{ x+\theta_t-\theta_*}{uy}\\ 
    uy\lb x-\theta_t-\theta_*\rb  & x-\theta_*\end{array}\rb,\\
   B_0=&\,\lb\begin{array}{cc} 0 & \frac{x\lb y-1\rb-\theta_0y-\theta_t+\theta_*}{tuy} \\ 
   \frac{\lb x\lb y-1\rb -\lb\theta_*+\theta_t\rb y-\theta_0\rb u}{t}  & 0 \end{array}\rb.
  \end{align*}
  The equation (\ref{linsys}) is a canonical form of rank 2 linear system with two regular singular points ($z=0,1$) and one irregular singularity of Poincaré rank $1$ ($z=\infty$). The 2nd equation describes monodromy preserving deformations of this system.
  The eigenvalues of $A_0$ and $A_1$ are given respectively by $\pm \theta_0$ and $\pm \theta_t$ and correspond to local monodromy at $0$ and $1$, whereas $\pm\theta_{*}$ are the exponents of formal monodromy at $z=\infty$.
  
  The compatibility condition $\partial_t A-\partial_z B=\left[A,B\right]$ gives a system of 1st order ODEs satisfied by $x$, $y$, $u$:
  \begin{align*}
  & t\dot x=\lb x-\theta_0\rb\lb x-\theta_t-\theta_*\rb y-
  \lb x+\theta_0\rb\lb x+\theta_t-\theta_*\rb y^{-1},\\
  & t\dot y=\lb\theta_0+\theta_t+\theta_*-2x\rb\lb y-1\rb^2+
  2\lb \theta_0+\theta_t\rb\lb y-1\rb+ty,\\
  & tu^{-1}\dot u=\lb x-\theta_t-\theta_*\rb y+\lb x+\theta_t-\theta_*\rb y^{-1}-2x.
  \end{align*}
  The first two equations are decoupled from the 3rd one. 
  It is thus straightforward to check that $y$ satisfies the standard form of the (non-degenerate) Painlevé V equation
  \ben
  \ddot y=\lb\frac{1}{2y}+\frac{1}{y-1}\rb \dot y^2-\frac{\dot y}{t}+\frac{\lb y-1\rb^2}{t^2}\left[\alpha y+\frac{\beta}{y}\right]+\frac{\gamma y}{t}+\frac{\delta y\lb y+1\rb}{y-1},
  \ebn
  with parameters 
  \ben
  \lb\alpha,\beta,\gamma,\delta\rb =\lb\frac{\lb\theta_0-\theta_t-\theta_*\rb^2}{2},-\frac{\lb\theta_0-\theta_t+\theta_*\rb^2}{2}, 1-2\theta_0-2\theta_t,-\frac12\rb.
  \ebn
  The value $\delta=-\frac12$ is a convenient normalization rather than a constraint since any non-zero value of $\delta$ can be achieved by a suitable rescaling of  $t$. The directions $\arg t=\frac{\pi k}2$ ($k=0,\ldots,3$) play a distinguished role in this normalization. For generic complex $\delta\ne 0$, these rays would rotate appropriately. The case $\delta=0$ where PV can be reduced to Painlevé III equation is left outside the scope of this note. Introducing
  \ben
  \begin{aligned}H\lb t\rb:=&\,-\frac12\operatorname{res}_{z=0}\operatorname{Tr} A^2\lb z,t\rb+
  \frac{\lb t-\theta_*\rb\theta_*}{2}=\\
  =&\,\lb x-\theta_0\rb\lb x-\theta_t-\theta_*\rb y +
  \frac{\lb x+\theta_0\rb\lb x+\theta_t-\theta_*\rb}{y}-
  \frac{\lb 2 x-\theta_*+t\rb \lb 2x-\theta_*\rb}{2},
  \end{aligned}
  \ebn
  it is easy to check that $H\lb t\rb$ satisfies the equation \eqref{sigpv}.\vspace{0.2cm}
  
  \noindent\textbf{3.1.2. Tau function}. Recall that, in the description of our results in Subsection~\ref{subsecintropv}, the PV tau function was related to the hamiltonian $H\lb t\rb$ by
  \beq\label{tauH}
  t\partial_t\ln\tau= H+\frac{\theta_*\lb t+\theta_*\rb}{2}.
  \eeq 
  This is slightly different from the original definition of the Jimbo-Miwa-Ueno tau function $\tau_{\mathrm{JMU}}\lb t\rb$.  In order to describe the difference, note that the system \eqref{linsys} has a unique \textit{formal} solution of the form
    \beq\label{formsolinf}
    \begin{gathered}
    \Phi^{\lb\infty\rb}_{\text{form}}\lb z\rb = e^{\Theta^{(\infty)}\lb z\rb} G^{\lb\infty\rb}_{\text{form}}\lb z\rb ,\\
    G^{\lb\infty\rb}_{\text{form}}\lb z\rb=\mathbb 1+\sum_{k=1}^{\infty}g^{\lb \infty\rb}_k z^{-k},\qquad \Theta^{(\infty)}\lb z\rb=\lb
      \frac{tz}2+\theta_*\ln z\rb\sigma_z.
    \end{gathered}
    \eeq
    The coefficients $g^{(\infty)}_k$ can be determined recursively; e.g. the 1st of them reads
    \beq\label{g1inf}
    g^{(\infty)}_1=\frac1t\lb\begin{array}{cc} 
    -H-\theta_0^2-\theta_t^2+\frac{\theta_*\lb\theta_*-t\rb}{2}& \frac{xy-x-\theta_0y+\theta_*-\theta_t}{uy}\\
    -u\lb xy-x-\lb\theta_*+\theta_t\rb y-\theta_0\rb &
    H+\theta_0^2+\theta_t^2-\frac{\theta_*\lb\theta_*-t\rb}{2}
    \end{array}\rb.
    \eeq
    The conventional definition of $\tau_{\mathrm{JMU}}\lb t\rb$ (see e.g. \cite[eq. (1.23)]{JMU}) for the system \eqref{linsys} thus reduces to
    \beq\label{jmucomp}
    \partial_t\ln\tau_{\mathrm{JMU}}=-\operatorname{res}_{z=\infty}
    \operatorname{Tr}\lb\partial_z G^{\lb\infty\rb}_{\text{form}} {G^{\lb\infty\rb}_{\text{form}}}^{-1} \partial_t\Theta^{(\infty)}\rb
    =-\frac12\operatorname{Tr}g_1^{(\infty)}\sigma_z.
    \eeq
    The last formula, together with \eqref{g1inf} and \eqref{tauH}, implies that
    \beq\label{taujmuvsour}
    \tau_{\mathrm{JMU}}\lb t\rb=t^{\theta_0^2+\theta_t^2-\theta_*^2}\tau\lb t\rb.
    \eeq
    The reason for using $\tau\lb t\rb$ instead of $\tau_{\mathrm{JMU}}\lb t\rb$ is that the former function may turn out to be more natural from the point of view of irregular CFT.\vspace{0.2cm}

   \noindent\textbf{3.1.3. Monodromy}. The canonical solutions $\Phi^{\lb\infty\rb}_{k}\lb z\rb$ are uniquely determined by the requirement to be asymptotic to $\Phi^{\lb\infty\rb}_{\text{form}}\lb z\rb$  inside the Stokes sectors at $z=\infty$,
  \ben\Omega_k=\left\{z\in\Cb : |z|>R,-\tfrac{\pi}{2}+\pi\lb k-2\rb <\arg z+\arg t<\tfrac{3\pi}{2}+\pi\lb k-2\rb\right\},\qquad  
  k\in\mathbb N.
  \ebn
  These solutions satisfy $\Phi^{\lb\infty\rb}_{k+1}\lb z\rb=
  S_k\Phi^{\lb\infty\rb}_{k}\lb z\rb $, where $S_{2k}=\lb\begin{array}{cc} 1 & 0 \\ s_{2k} & 1 \end{array}\rb$, $S_{2k+1}=\lb\begin{array}{cc} 1 & s_{2k+1} \\ 0 & 1 \end{array}\rb$ are called the Stokes matrices. Also, $\Phi^{\lb\infty\rb}_{k+2}\lb z e^{2\pi i}\rb=e^{2\pi i \theta_*\sigma_z}\Phi^{\lb\infty\rb}_{k}\lb z\rb $, which implies that all Stokes matrices can be expressed in terms of $S_1$ and $S_2$. 
    
   Let us set $\Phi^{\lb\infty\rb}\lb z\rb\equiv \Phi^{\lb\infty\rb}_{2}\lb z\rb$ and fix the branch of $\Phi^{\lb\infty\rb}$ by introducing a branch cut
  along $\mathbb R_{\ge0}$. The monodromy matrix 
  around $\infty$ defined by
  $\Phi^{\lb\infty\rb}\bigl( z e^{-2\pi i}\bigr)=M_{\infty} \Phi^{\lb\infty\rb}\lb z\rb $ may be expressed as
  \beq\label{minf}
  M_{\infty}=S_1 e^{-2\pi i \theta_*\sigma_z}S_2.
  \eeq
  Denote by $M_0,M_t$ the anticlockwise monodromies of $\Phi^{\lb\infty\rb}\lb z\rb$ around $0$ and $1$. Together with $M_{\infty}$, they satisfy a cyclic relation 
  $M_0M_t M_{\infty}=\mathbb 1$. 
  The matrices $M_0$, $M_t$, $M_{\infty}$, $S_1$, $S_2$ constitute the  set of generalized monodromy data for the linear system \eqref{linsys} with fixed local and formal monodromy exponents $\theta_0$, $\theta_t$, $\theta_*$. To parameterize these data in a convenient way, define
  \ben
  \begin{gathered}
  p_0:=\operatorname{Tr} M_0=2\cos2\pi\theta_0,\qquad
  p_t:=\operatorname{Tr} M_t=2\cos2\pi\theta_t,\qquad
  e_*:=e^{\pi i \theta_*},
  \end{gathered}
  \ebn
  and introduce $W\lb A_3\rb$-invariant combinations
  \ben
  A_{\pm}:=e_*^{\mp 1}p_0+e_*^{\pm 1}p_t,\qquad 
   A_*:=e_*^2+e_*^{-2}+p_0 p_t. 
  \ebn
  Thanks to (\ref{minf}), 
  $M_0$, $M_t$ may be parameterized as
  \beq\label{M0ett}
  M_0=\lb\begin{array}{ll}
  X_+ &  W\\ V & X_-
  \end{array}\rb e_*^{\sigma_z}S_2^{-1},\qquad 
  M_t= S_1^{-1}e_*^{\sigma_z} \lb\begin{array}{rr}
    X_- &  -W\\ -V & X_+
    \end{array}\rb,\qquad X_+X_--VW=1.
  \eeq
  Also, introduce $X_{\sigma}:=\operatorname{Tr} M_t M_0=\operatorname{Tr} M_{\infty}$ and notice that $X_{\sigma}=e_*^{-2}+e_*^2\lb 1+s_1s_2\rb$.
   
  Assuming that $s_1s_2\ne0$ and using the relations $p_{\nu}=\operatorname{Tr}M_{\nu}$ ($\nu=0,t$), the coefficients $V$, $W$ can be expressed in terms of $X_{\pm}$ and $s_{1,2}$:
  \beq\label{tracesp0t}
  V=\lb p_t-e_*^{-1} X_+-e_* X_-\rb e_*^{-1}s_1^{-1},\qquad 
  W=\lb e_* X_++e_*^{-1} X_--p_0\rb e_*^{-1}s_2^{-1}. 
  \eeq  
  For any $\kappa\neq 0$, the  matrix $\tilde\Phi^{\lb\infty\rb}\lb z\rb=\kappa^{\frac{\sigma_z}2}\Phi^{\lb\infty\rb}\lb z\rb \kappa^{-\frac{\sigma_z}2}$ has asymptotic expansions of the form \eqref{formsolinf}, and the only change in the linear system amounts to a rescaling $u\mapsto \tilde u=\kappa^{-1}u$. We may therefore identify the corresponding monodromy and Stokes matrices, whose parameters are obtained by rescalings
  \ben
  \lb X_{\sigma},X_+,X_-,V,W,s_1,s_2\rb\mapsto \lb X_{\sigma},X_+,X_-,\kappa^{-1} V,\kappa W,\kappa s_1,\kappa^{-1}s_2\rb. 
  \ebn
  This in turn identifies the space of Painlevé V monodromy data with a degenerate affine cubic surface in $\mathbb C^3$ expressing the unit determinant condition
  $X_+X_--VW=1$:
  \beq\label{MPV}
  \mathcal M:=\bigl\{\lb X_{\sigma},X_+,X_-\rb\in\mathbb C^3\,\bigl|\, X_{\sigma}X_+X_-+X_+^2+X_-^2-X_{\sigma}-A_+ X_+-A_- X_-+A_*=0\bigr\}.
  \eeq
   A pair $\lb X_+,X_-\rb\in\mathbb C^2$ thereby determines the monodromy up to simultaneous diagonal conjugation. 
  \begin{rmk}The cubic can also be rewritten in the form
  \beq\label{MPV2}
  \lb X_{\sigma}-e_*^2-e_*^{-2}\rb \lb X_+X_--1\rb+\lb e_* X_++e_*^{-1} X_--p_0\rb
  \lb e_*^{-1} X_++e_* X_--p_t\rb =0,
  \eeq
  which allows to relax the condition $s_1s_2\ne0$. Indeed, for $s_1s_2=0$ we have $X_{\sigma}= e_*^2+e_*^{-2}$ so that the cubic splits
  into a union of two lines,  $e_* X_++e_*^{-1} X_-=p_0$ and $e_*^{-1} X_++e_* X_-=p_t$. The same equations follow from the computation of $\operatorname{Tr}M_0$, $\operatorname{Tr}M_t$,
  cf \eqref{tracesp0t}.
  \end{rmk}
   
  For fixed $X_{\sigma}= 2\cos2\pi\sigma$, the resulting conic in 
  the variables $X_{\pm}$ admits a rational parameterization. The result for generic monodromy may be written in the form \eqref{zetap}, where $e^{2\pi i\eta}$ plays the role of a rational parameter. Its geometric significance will become more transparent in the Riemann-Hilbert setting. For now, let us simply note that parameterizing the monodromy by the pair $\lb\sigma,\eta\rb$ instead of $\lb X_{+},X_-\rb$ is particularly adapted for  asymptotic analysis of the Painlevé V tau function as $t\to0$, cf Theorem~\ref{theoB}. The inverse formulas are given by
  \begin{subequations}\label{mmpar}
  \begin{align}\label{sigmafromXpm}
  \cos2\pi\sigma=&\,\frac{\lb e^{i\pi\theta_*} X_++e^{-i\pi\theta_*} X_--2\cos2\pi\theta_0\rb
    \lb e^{-i\pi\theta_*} X_++e^{i\pi\theta_*} X_--2\cos2\pi\theta_t\rb}{2\lb 1-X_+X_-\rb}+\cos2\pi\theta_*,\\
  e^{2\pi i \eta}=&\,\frac{\sin2\pi\sigma\lb e^{-i\pi\sigma}X_++e^{i\pi\sigma}X_-\rb
  +2\cos2\pi\theta_0\sin\pi\lb\theta_*-\sigma\rb
  -2\cos2\pi\theta_t\sin\pi\lb\theta_*+\sigma\rb}{4
  \sin\pi\lb \theta_0+\theta_t-\sigma\rb
        \sin\pi\lb \theta_t-\theta_0-\sigma\rb
        \sin\pi\lb \theta_*-\sigma\rb}.
  \end{align}
  \end{subequations}
  The 1st of them is nothing but \eqref{MPV2} in a more explicit form, while the 2nd is a suitable linear combination of equations \eqref{zetap}. The solution of \eqref{sigmafromXpm} for $\sigma$ can be chosen arbitrarily: the periodicity of the expansion \eqref{fourier0} with respect to the integer shifts $\sigma\mapsto\sigma\pm1$ and invariance of conformal blocks
  $\mathcal B\lb \theta_*;\sigma;\substack{\theta_t \\ \theta_0};t\rb $  with respect to the sign flip $\sigma\mapsto -\sigma$ automatically takes care of this ambiguity.

  \subsection{Riemann-Hilbert problem}
\begin{figure}[h!]\centering
    \begin{tikzpicture}[scale=0.6]
    \draw  (0,0) circle (1);
    \draw  (3.5,0) circle (1);
    \draw  (0,0) circle (6.5);
    \draw  (0,0) circle (8.5);
    \draw  [dashed,color=red](0,0) circle (7.3);
  \draw  [->,>=latex,color=red] (0,7.3) -- (-0.2,7.3);   
    \draw  (1,0) -- (2.5,0);
    \draw  (4.5,0) -- (6.5,0);
    \draw  (8.5,0) -- (12,0);
    \draw  (-12,0) -- (-8.5,0);
    \draw  [->,>=latex] (2.04,0) -- (2.06,0);
    \draw  [->,>=latex] (0,1) -- (-0.2,1);
    \draw  [->,>=latex] (0,-1) -- (0.2,-1);
    \draw  [->,>=latex] (3.5,1) -- (3.3,1);
    \draw  [->,>=latex] (10.3,0) -- (10.4,0);
    \draw  [->,>=latex] (-10.3,0) -- (-10.4,0);
    \draw  [->,>=latex] (5.3,0) -- (5.6,0); 
    \draw  [->,>=latex] (-0.2,6.5) -- (0.2,6.5); 
    \draw  [->,>=latex] (0.2,8.5) -- (-0.2,8.5);   
     \draw  [->,>=latex] (0.2,-6.5) -- (-0.2,-6.5);    
     \draw  [->,>=latex] (-0.2,-8.5) -- (0.2,-8.5);    
    \draw  [fill] (0,0) circle (0.05);
    \draw  [fill] (3.5,0) circle (0.05); 
    \draw  (0.05,0.05) node[below left] {\small $0$}; 
    \draw  (3.55,0.05) node[below left] {\small $1$}; 
    \draw  (1.7,0) node[below] {\small $M_0^{-1}$}; 
    \draw  (5.5,-0.1) node[below] {\small $M_{\infty}$};
    \draw  (-1,0) node[left] {\small $\lb -z\rb^{-\Theta_0}C_0^{-1}$};
    \draw  (3.5,1) node[above] {\small $\lb 1-z\rb^{-\Theta_t}C_{1,+}^{-1}$}; 
    \draw  (3.5,-1) node[below] {\small $\lb 1-z\rb^{-\Theta_t}C_{1,-}^{-1}$};     
    \draw  (0.35,0.4) node[draw,circle,fill=Gray!40] {\tiny $\,$I$\,$};   
    \draw  (3.8,0.45) node[draw,circle,fill=Gray!40] {\tiny II}; 
    \draw  (-3,3) node[draw,circle,fill=Gray!40] {\tiny III};   
    \draw  (-5.6,5.6) node[draw,circle,fill=Gray!40] {\tiny IV};  
    \draw  (-10.6,5.6) node[draw,circle,fill=Gray!40] {\tiny V};   
    \draw  (-10.6,-3.6) node[draw,circle,fill=Gray!40] {\tiny VI};
    \draw  (-10.6,0.1) node[above] {$S_2$};  
    \draw  (10.6,0.1) node[above] {$S_1 e^{-2\pi i \theta_*\sigma_z}$}; 
    \draw  (6.3,7.1) node[above] {$\lb -z\rb^{-\mathfrak S}U^{-1}$};  
    \draw  (0.3,5.1) node[above] {$\lb -z\rb^{-\mathfrak S}U^{-1}$}; 
    \draw  (6.8,-8) node[above] {$\lb -z\rb^{-\mathfrak S}U^{-1}S_2^{-1}$};          
    \end{tikzpicture}
             \caption{\label{RHPcont}
             Transformation of the RHP contour.}
    \end{figure}
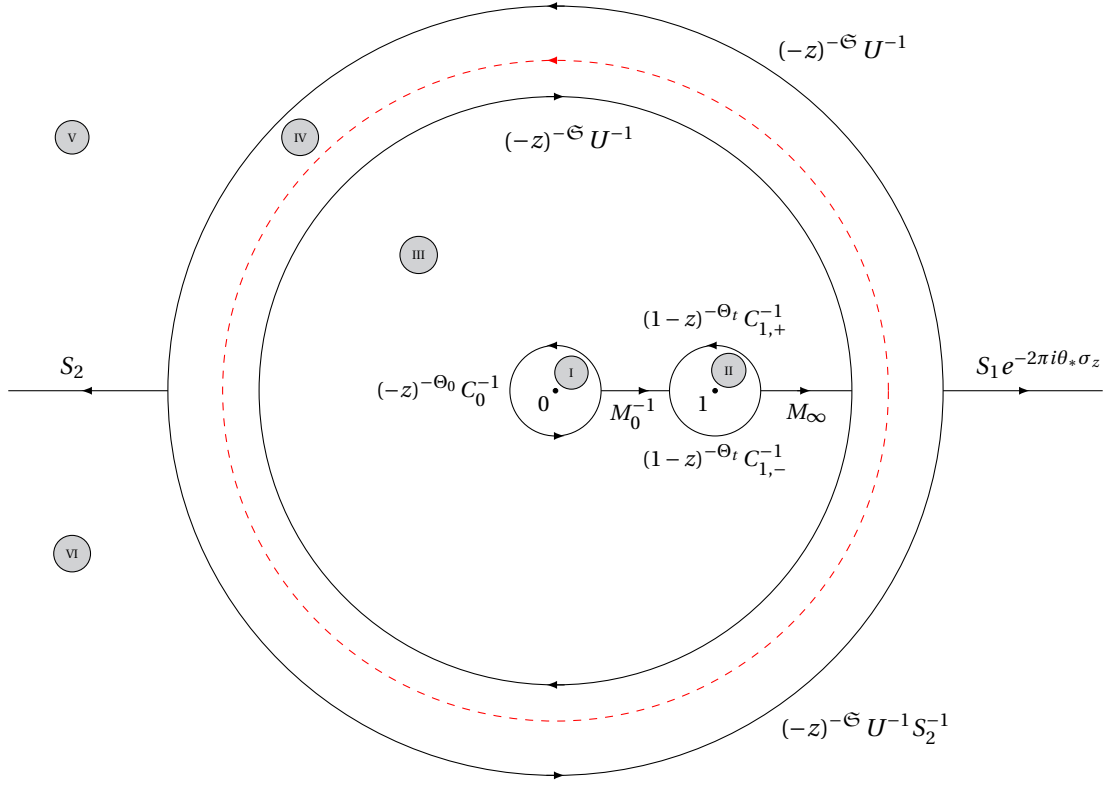

  Let us now transform the linear equation \eqref{linsys} into a suitable Riemann-Hilbert problem (RHP) following the method developed in \cite{GL16,GL17,CGL}. We start from the contour $\hat{\Gamma}$ depicted in Fig.~\ref{RHPcont} where it is assumed for simplicity that $\operatorname{arg}t=0$. We also assume that $M_{\infty}$ is diagonalizable and introduce the notation
  \beq\label{diagmi}
  M_{\infty}=Ue^{-2\pi i \mathfrak S}U^{-1},\qquad \mathfrak S=\operatorname{diag}\left\{\sigma,-\sigma\right\},
  \qquad \Theta_0=\operatorname{diag}\left\{\theta_0,-\theta_0\right\},\qquad
  \Theta_t=\operatorname{diag}\left\{\theta_t,-\theta_t\right\}.
  \eeq
  The solution $\hat{\Psi}\lb z\rb$ of this RHP is related to that of \eqref{linsys} via the identification
  \ben
  \hat{\Psi}\lb z\rb=\begin{cases}
  \lb -z\rb^{-\Theta_0}C_0^{-1}\Phi^{(\infty)}\lb z\rb ,\quad & z\in D_{\mathrm{I}},\\
  \lb 1-z\rb^{-\Theta_t}C_{1,\pm}^{-1}\Phi^{(\infty)}\lb z\rb ,\quad & z\in D_{\mathrm{II}},\Im z\gtrless 0,\\  
  \Phi^{(\infty)}\lb z\rb, & z\in D_{\mathrm{III}}\cup D_{\mathrm{V}},\\
  \lb -z\rb^{-\mathfrak S} U^{-1}\Phi^{(\infty)}\lb z\rb, & z\in D_{\mathrm{IV}},\\  
  \Phi^{(\infty)}_3\lb z\rb, & z\in D_{\mathrm{VI}},
  \end{cases}
  \ebn
  Let $\mathcal C$ be the circle indicated in Fig.~\ref{RHPcont} by a dashed red line. Denote by $\Psi_+\lb z\rb$ and $\Psi_-\lb z\rb$ the solutions of the auxiliary RHPs posed on $\hat\Gamma_+$ and $\hat\Gamma_-$, the connected components of $\hat\Gamma$ located outside and inside $\mathcal C$, respectively, and characterized by the same jumps. The existence of such solutions follows from their explicit construction below. 
  
  At the next step, introduce
  \ben
  \bar{\Psi}\lb z\rb= \begin{cases}
  \Psi_{+}^{-1}\lb z\rb \hat{\Psi}\lb z\rb,\qquad z\text{ outside }\mathcal C,\\
  \Psi_{-}^{-1}\lb z\rb \hat{\Psi}\lb z\rb,\qquad z\text{ inside }\mathcal C.
  \end{cases}
  \ebn
  The function $\bar{\Psi}\lb z\rb$ thus has no jumps except on $\mathcal C$, where the jump matrix is given by 
  \ben
  J\lb z\rb=\bar\Psi_+\lb z\rb \bar\Psi_-^{-1}\lb z\rb=\Psi_-^{-1}\lb z\rb \Psi_+\lb z\rb.
  \ebn
  This brings us to the general setup of \cite{CGL} for one circle, with $\Psi_{\pm}$ and $\bar\Psi_{\pm}$ providing the direct and dual factorization of $J$. As the notation suggests, the indices $\pm$ refer to $\mathrm{GL}_2\lb \Cb\rb $-functions analytic inside/outside $\mathcal C$. 
  \subsection{Tau function as Fredholm determinant}
  \begin{defin} Let $H=L^2\lb \mathcal C,\Cb^2\rb$ with canonical decomposition $H=H_+\oplus H_-$.
  Let $\mathsf a:H_-\to H_+$ and $\mathsf d:H_+\to H_-$ be the integral operators
  \ben
  \lb \mathsf a f\rb\lb z\rb=\frac{1}{2\pi i }\int_{\mathcal C}\mathsf a\lb z,z'\rb f\lb z'\rb dz',\qquad 
 \lb \mathsf d f\rb\lb z\rb=\frac{1}{2\pi i }\int_{\mathcal C}\mathsf d\lb z,z'\rb f\lb z'\rb dz',  
  \ebn
  whose integrable kernels are given by
  \beq\label{adkers}
  \mathsf a\lb z,z'\rb=\frac{\mathbb 1-\Psi_+\lb z\rb\Psi_+^{-1}\lb z'\rb}{z-z'},\qquad 
  \mathsf d\lb z,z'\rb=\frac{\Psi_-\lb z\rb\Psi_-^{-1}\lb z'\rb-\mathbb 1}{z-z'}.  
  \eeq
  The tau function $\tau\left[J\right]$ associated to the Riemann-Hilbert problem described by the pair $\lb \mathcal C,J\rb$ is defined as the Fredholm determinant
  \beq\label{taufredd}
  \tau\left[J\right]=\operatorname{det}_H\lb\begin{array}{cc}
  \mathbb 1 & \mathsf a \\ \mathsf d & \mathbb 1\end{array}\rb.
  \eeq
  \end{defin}
  
  The relation of $\tau\left[J\right]$ to the Painlev\'e V tau function (associated to the monodromy data parameterized as above) is provided by the following result.
  \begin{prop} The Painlevé V tau function (\ref{tauH})  can be expressed as
  \beq\label{tautauJ}
  \tau\lb t\rb=\operatorname{const}\cdot t^{\sigma^2-\theta_0^2-\theta_t^2}\tau\left[J\right].
  \eeq
  \end{prop}
  \pf Essentially the same as the proof of Corollary 2.7 in \cite{CGL}, making use of the Widom's differentiation formula \cite{Widom1,IJK}. The PV analog of the topological decompositions (2.35) from \cite{CGL} takes the form\footnote{The bordered cusped Riemann surfaces were associated to Painlevé monodromy manifolds in \cite{CM,CMR}.}
     \beq\label{taugraph}
     \begin{gathered}
     \tau_{\mathrm{JMU}}\lb\;
     \vcenter{\hbox{\includegraphics[height=5.5ex]{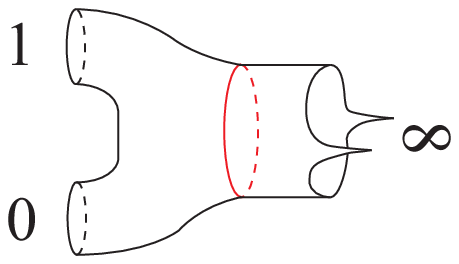}}}\;\rb= 
     \tau_{\mathrm{JMU}}\lb\;
       \vcenter{\hbox{\includegraphics[height=5ex]{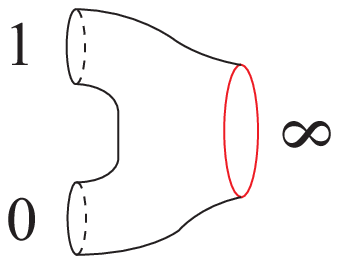}}}\;\rb
       \tau_{\mathrm{JMU}}\lb\;
        \vcenter{\hbox{\includegraphics[height=4ex]{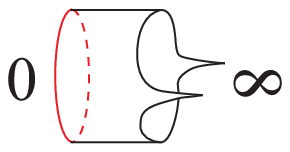}}}\;\rb\;
     \det\lb\begin{array}{cc}
     \mathbb 1 & \mathsf a\lb\;
       \vcenter{\hbox{\includegraphics[height=4ex]{LNRpantsR.eps}}}\;\rb
      \\ 
       \mathsf d\lb\;
          \vcenter{\hbox{\includegraphics[height=5ex]{LNRpantsL.eps}}}\;\rb & \mathbb 1
     \end{array}\rb .
       \end{gathered}
     \eeq
  The left side of this relation represents the Jimbo-Miwa-Ueno tau function of the linear system \eqref{linsys} with 2 regular and 1 irregular singular point, related to $\tau\lb t\rb$ by \eqref{taujmuvsour}. The determinant on the right is nothing but $\tau\left[J\right]$. The remaining two factors are the tau functions of auxiliary linear systems whose solutions are related to $\Psi_-\lb z\rb$ and $\Psi_+\lb z\rb$. The first of these systems has 3 regular singularities and the second has 1 regular + 1 irregular singular point. In the first case, the tau function  does not depend on $t$ and may be omitted. Therefore, it remains to compute
  $\tau^{(+)}\lb t\rb\equiv \tau_{\mathrm{JMU}}\lb\;
          \vcenter{\hbox{\includegraphics[height=4ex]{LNRpantsR.eps}}}\;\rb $.
  
  The relevant auxiliary linear problem has the form
  \beq\label{linsysR}
  \partial_z\Phi_+=\Phi_+\lb \frac{t\sigma_z}{2}+\frac1z\lb
  \begin{array}{cc}\theta_* & r_+\lb\sigma-\theta_*\rb \\ 
  r_+^{-1}\lb \sigma+\theta_*\rb & -\theta_*\end{array}\rb\rb,
  \eeq
  which follows from the known exponential behavior and known exponents $\sigma$, $\theta_*$ of local monodromy around $0$ and formal monodromy around $\infty$. The formal solution of the system \eqref{linsysR} at $\infty$ is given by the series analogous to 
  \eqref{formsolinf},
      \beq\label{formsolinfR}
      \begin{gathered}
      \Phi^{\lb\infty\rb}_{+,\text{form}}\lb z\rb = e^{\Theta^{(\infty)}\lb z\rb} \left[\mathbb 1+\sum\nolimits_{k=1}^{\infty}g^{\lb \infty\rb}_{+,k} z^{-k}\right],
      \end{gathered}
      \eeq
  with the same $\Theta^{(\infty)}\lb z\rb=\lb \frac{tz}2+\theta_*\ln z\rb\sigma_z$. For the computation of the tau function, we need the 1st coefficient of this series, which is given by \ben
  g_{+,1}^{(\infty)}=\frac1t\lb\begin{array}{cc}
  \theta_*^2-\sigma^2 & r_+\lb \sigma-\theta_*\rb \\
  -r_+^{-1}\lb\sigma+\theta_*\rb & \sigma^2-\theta_*^2\end{array}\rb.
  \ebn
  Similarly to \eqref{jmucomp}, we have $\partial_t\tau^{(+)}=-\frac12\operatorname{Tr}g_{+,1}^{(\infty)}\sigma_z=\lb\sigma^2-\theta_*^2\rb/t$. This implies that $\tau^{(+)}\lb t\rb=\operatorname{const}\cdot t^{\sigma^2-\theta_*^2}$ and, in combination with \eqref{taugraph} and \eqref{taujmuvsour}, yields the result \eqref{tautauJ}.
  \epf
  
  In order to make the Fredholm determinant \eqref{taufredd} completely explicit, we need  expressions for the parametrices $\Psi_{\pm}\lb z\rb$ for $z\in D_{\mathrm{IV}}$, i.e. inside the annulus containing $\mathcal C$. To construct $\Psi_+\lb z\rb$, consider the following fundamental solutions of \eqref{linsysR}:
  \begin{subequations}
  \beq
   \Phi_+^{(0)}\lb z\rb=
   t^{-\mathfrak S}
   \lb\begin{array}{cc}
   M_{-\theta_*,-\frac12+\sigma}\lb t z\rb & 
   \frac{\theta_*-\sigma}{2\sigma\lb 2\sigma+1\rb}\,M_{-\theta_*,\frac12+\sigma}\lb t z\rb \\
   \frac{\theta_*+\sigma}{2\sigma\lb 2\sigma-1\rb}\,M_{-\theta_*,\frac12-\sigma}\lb t z\rb & M_{-\theta_*,-\frac12-\sigma}\lb t z\rb
   \end{array}\rb
   \lb\begin{array}{cc}
   r_+^{-\frac12} & r_+^{\frac12} \frac{\sigma-\theta_*}{\sigma+\theta_*}\\
   -r_+^{-\frac12} & r_+^{\frac12}
   \end{array}\rb,
  \eeq
  \beq
  \Phi_+^{(\infty)}\lb z\rb=\frac{1}{\sqrt z}
  \lb\begin{array}{cc}
  \lb e^{-i\pi}t\rb^{-\frac12-\theta_*} & 0 \\ 0 & t^{-\frac12+\theta_*}
  \end{array}\rb
  \lb\begin{array}{cc}
  W_{\frac12+ \theta_*,\sigma}\lb e^{-i\pi}t z\rb & 
  r_+\lb\theta_*-\sigma\rb W_{-\frac12+ \theta_*,\sigma}\lb e^{-i\pi}t z\rb \\
  -r_+^{-1}\lb\theta_*+\sigma\rb W_{-\frac12- \theta_*,\sigma}\lb t z\rb & 
  W_{\frac12- \theta_*,\sigma}\lb t z\rb
  \end{array}\rb,
  \eeq
  \end{subequations}
  where $M_{\kappa,\mu}\lb x\rb$, $W_{\kappa,\mu}\lb x\rb$ denote the Whittaker functions.
  
  The solution $\Phi_+^{(\infty)}\lb z\rb$ models the asymptotic behavior and monodromy of $\Phi^{(\infty)}_2\lb z\rb$ at $z=\infty$; in particular,
  \ben
  \Phi_+^{(\infty)}\lb z\to\infty\rb\simeq e^{\Theta^{(\infty)}\lb z\rb}\left[\mathbb 1+O\lb z^{-1}\rb\right],\qquad \arg tz\in\lb-\tfrac{\pi}{2},\tfrac{3\pi}{2}\rb.
  \ebn
  The formulas for canonical solutions of \eqref{linsysR} in other sectors are straightforward to obtain: it suffices to add appropriate powers of $e^{2\pi i }$ to the arguments of Whittaker functions  and adapt the diagonal normalization prefactor accordingly. The Stokes multipliers $s_{1,+}$, $s_{2,+}$ for $\Phi_+^{(\infty)}\lb z\rb$ can then be found from the analytic continuation formulae for $W_{\kappa,\mu}\lb x\rb$. Identifying them with $s_1$, $s_2$, one finds that
  \ben
  s_{1}=-\frac{2\pi i r_+ t^{-2\theta_*}}{\Gamma\lb\sigma-\theta_*\rb \Gamma\lb 1-\sigma-\theta_*\rb},\qquad s_{2}=-\frac{2\pi i r_+^{-1}t^{2\theta_*}e^{-2\pi i \theta_*}}{\Gamma\lb\theta_*+\sigma\rb\Gamma\lb 1+\theta_*-\sigma\rb}.
  \ebn
  The latter equations should be viewed as determining the parameter $r_+$ in \eqref{linsysR} in terms of monodromy data at $\infty$ that we want to reproduce. One may verify, for instance, that $\operatorname{Tr}\lb S_1e^{-2\pi i \theta_*\sigma_z}S_2\rb=2\cos2\pi\sigma$, as expected.
  
  The solutions $\Phi_+^{(0)}\lb z\rb$ and $\Phi_+^{(\infty)}\lb z\rb$ are related by 
  \ben
  \Phi_+^{(\infty)}\lb z\rb=Ut^{\mathfrak S}\Phi_+^{(0)}\lb z\rb, 
  \ebn 
  \ben
  U=\lb\begin{array}{cc}
  \ds\frac{\Gamma\lb-2\sigma\rb e^{i\pi\lb\theta_*-\sigma\rb} r_+^{\frac12}t^{-\theta_*}}{\Gamma\lb-\theta_*-\sigma\rb} &
  -\ds\frac{\Gamma\lb 2\sigma\rb e^{i\pi\lb\theta_*+\sigma\rb} r_+^{\frac12}t^{-\theta_*}}{\Gamma\lb-\theta_*+\sigma\rb}\vspace{0.1cm}\\
  \ds -\frac{\lb\theta_*+\sigma\rb\Gamma\lb -2\sigma\rb r_+^{-\frac12}t^{\theta_*}}{
    \Gamma\lb 1+\theta_*-\sigma\rb} & \ds \frac{\Gamma\lb 2\sigma\rb r_+^{-\frac12}t^{\theta_*}}{
  \Gamma\lb \theta_*+\sigma\rb}
  \end{array}\rb.
  \ebn
  Since $\Phi_+^{(0)}\lb z\rb$ has diagonal monodromy around $z=0$, it should not appear as a surprise that the connection matrix $U$ ideed diagonalizes $M_{\infty}$ as in \eqref{diagmi}. Observe that, when expressed in terms of $s_{1}$, $s_2$ instead of $r_+$,  $U$ becomes indepenent of $t$.
  
  Therefore, the solution of the RHP for $\Psi_+\lb z\rb$ inside $D_{\mathrm{IV}}$ can be chosen as
  \beq\label{psipz}
  \Psi_+\lb z\rb = e^{i\pi\mathfrak S}t^{\mathfrak S}
     \lb\begin{array}{cc}
     \lb tz\rb^{-\sigma} M_{-\theta_*,-\frac12+\sigma}\lb t z\rb & 
     \frac{\theta_*-\sigma}{2\sigma\lb 2\sigma+1\rb}\,\lb tz\rb^{-\sigma}M_{-\theta_*,\frac12+\sigma}\lb t z\rb \\
     \frac{\theta_*+\sigma}{2\sigma\lb 2\sigma-1\rb}\, \lb tz\rb^{\sigma} M_{-\theta_*,\frac12-\sigma}\lb t z\rb & \lb tz\rb^{\sigma}M_{-\theta_*,-\frac12-\sigma}\lb t z\rb
     \end{array}\rb,
  \eeq
  which is nothing but suitably normalized $\Phi_+^{(0)}\lb z\rb$ with singular behavior near $z=0$ factored out. The choice of normalization (multiplication by a constant matrix on the right) does not influence the form of the integral kernel $\mathsf{a}\lb z,z'\rb$ in \eqref{taufredd}.
  
  In a similar manner, $\Psi_-\lb z\rb$ can be related to the solution of an auxiliary linear system with regular singularities at $0$, $1$, $\infty$ characterized by local monodromy exponents $\pm \theta_0$, $\pm \theta_t$, $\pm \sigma$. These data determine the system up to a constant gauge transformation (irrelevant to our purposes). The solution can then be given in terms of Gauss hypergeometric functions
  $_2F_1\lb a,b,c,z\rb$. Consider, for instance,
  \ben
  \partial_z\Phi_-=\Phi_-\lb\frac{A_{-,0}}{z}+
  \frac{A_{-,1}}{z-1}\rb, 
  \ebn
  with
  \ben
  A_{-,1}=\frac1{2\sigma}\lb\begin{array}{cc}
  \sigma^2+\theta_t^2-\theta_0^2 & 
  \lb\sigma+\theta_t\rb^2-\theta_0^2 \vspace{0.1cm}\\
  \theta_0^2 -\lb\sigma-\theta_t\rb^2 &  \theta_0^2-\theta_t^2-\sigma^2\end{array}\rb, \qquad A_{-,0}=\mathfrak S-A_{-,1}.
  \ebn
  The solution of this system can be chosen as
  \beq
  \Phi_-^{(\infty)}\lb z\rb=\lb -z\rb^{\mathfrak S}\lb 1-z^{-1}\rb^{\theta_t}\lb\begin{array}{cc}
  \phi_{11}\lb z\rb & \phi_{12}\lb z\rb \\
  \phi_{21}\lb z\rb & \phi_{22}\lb z\rb
  \end{array}\rb,
  \eeq
  \beq\label{phialbe}
  \begin{aligned}
  \phi_{11}\lb z\rb=&\, _2F_1\lb\theta_t+\theta_0-\sigma,\theta_t-\theta_0-\sigma,-2\sigma,z^{-1}\rb, \\
  \phi_{12}\lb z\rb=&\,\frac{\lb \theta_t+\sigma\rb^2-\theta_0^2}{2\sigma\lb 2\sigma-1\rb} z^{-1}{}_2F_1\lb1+\theta_t+\theta_0-\sigma,1+\theta_t-\theta_0-\sigma,2-2\sigma,z^{-1}\rb,\\
  \phi_{21}\lb z\rb=&\,\frac{\lb \theta_t-\sigma\rb^2-\theta_0^2}{2\sigma\lb 2\sigma+1\rb} z^{-1}{}_2F_1\lb1+\theta_t+\theta_0+\sigma,1+\theta_t-\theta_0+\sigma,2+2\sigma,z^{-1}\rb,\\
  \phi_{22}\lb z\rb=&\, {}_2F_1\lb \theta_t+\theta_0+\sigma,\theta_t-\theta_0+\sigma,2\sigma,z^{-1}\rb. 
  \end{aligned}
  \eeq
  Since $ \Phi_-^{(\infty)}\lb z\rb$ has diagonal monodromy at $z=\infty$, any solution with monodromy $M_{\infty}$ and the same local exponents can be written (up to normalization) as $Us_-^{\sigma_z}\Phi_-^{(\infty)}\lb z\rb$ with some $s_-\in\mathbb C^*$. Accordingly, $\Psi_-\lb z\rb$ inside $D_{\mathrm{IV}}$ is given by
  \beq\label{psimz}
  \Psi_-\lb z\rb=s_-^{\sigma_z}\lb 1-z^{-1}\rb^{\theta_t}\lb\begin{array}{cc}
    \phi_{11}\lb z\rb & \phi_{12}\lb z\rb \\
    \phi_{21}\lb z\rb & \phi_{22}\lb z\rb
    \end{array}\rb,
  \eeq
  which allows to compute the explicit form of the integral kernel $\mathsf{d}\lb z,z'\rb$. To relate the parameter $s_-$ to the monodromy data, observe that the counterclockwise monodromy of $Us_-^{\sigma_z}\Phi_-^{(\infty)}\lb z\rb$ around $z=0$ is given by
  \beq\label{M0second}
  M_0=Us_-^{\sigma_z}
  C_{0\infty}e^{2\pi i \Theta_0} C_{0\infty}^{-1}
  s_-^{-\sigma_z}U^{-1},
  \eeq
    \beq
    C_{0\infty}=\left(\begin{array}{cc}
    \displaystyle \frac{
    \Gamma\left(-2\theta_0\right)
    \Gamma\left(-2\sigma\right)}{
    \Gamma\left(-\theta_0-\theta_t-\sigma\right)
    \Gamma\left(1-\theta_0+\theta_t-\sigma\right)} & 
  \displaystyle \frac{\Gamma\left(2\theta_0\right)
     \Gamma\left(-2\sigma\right)}{
     \Gamma\left(\theta_0-\theta_t-\sigma\right)
     \Gamma\left(1+\theta_0+\theta_t-\sigma\right)}    
     \vspace{0.2cm}\\
        \displaystyle \frac{\Gamma\left(-2\theta_0\right)
         \Gamma\left(2\sigma\right)}{
         \Gamma\left(-\theta_0-\theta_t+\sigma\right)
         \Gamma\left(1-\theta_0+\theta_t+\sigma\right)}  &
   \displaystyle \frac{\Gamma\left(2\theta_0\right)
      \Gamma\left(2\sigma\right)}{
      \Gamma\left(\theta_0-\theta_t+\sigma\right)
      \Gamma\left(1+\theta_0+\theta_t+\sigma\right)}   
    \end{array}\right),
    \eeq
  see e.g. \cite[Subsection~3.6]{ILP}. Comparing the monodromy matrix \eqref{M0second} with the parameterization \eqref{M0ett}, we deduce the identification
  \beq\label{etapar}
  s_-^{-2}=\frac{\Gamma^2\lb1-2\sigma\rb
  \Gamma\lb1+\theta_*+\sigma\rb
  \Gamma\lb 1+\theta_0+\theta_t+\sigma\rb
  \Gamma\lb 1-\theta_0+\theta_t+\sigma\rb}{
  \Gamma^2\lb1+2\sigma\rb\Gamma\lb1+\theta_*-\sigma\rb
  \Gamma\lb 1+\theta_0+\theta_t-\sigma\rb \Gamma\lb 1-\theta_0+\theta_t-\sigma\rb}\,
  e^{2\pi i \lb\eta-\sigma\rb}.
  \eeq
  The parameter $\eta$ thus has the meaning of ``relative twist'' of conjugacy classes of monodromy data associated to auxiliary linear problems; it describes how their solutions should be glued to reproduce the jumps of the original RHP.
 
  The results of this subsection may be summarized as follows. The Painlevé V tau function associated to monodromy described by the pair $\lb\sigma,\eta\rb$ as in \eqref{mmpar} admits Fredholm determinant representation \eqref{taufredd}--\eqref{tautauJ}. The integral kernels \eqref{adkers} are expressed in terms of the parametrices $\Psi_{\pm}\lb z\rb$, explicitly given by \eqref{psipz}, \eqref{psimz}, and $\phi_{ij}\lb z\rb$ and $s_-$ therein are given by \eqref{phialbe} and \eqref{etapar}.
  
  \subsection{Proof of Theorem B}
  The Fredholm determinant representation makes it straightforward to obtain the asymptotics of $\tau\left[J\right]$ as $t\to 0$. It suffices to write the integral kernels of $\mathsf a$ and $\mathsf d$ in the Fourier basis,
  \begin{align}
  \mathsf a\lb z,z'\rb=&\, \sum_{p,q\in\mathbb N+\frac12}\mathsf a_{p,-q}z^{p-\frac12}z'^{q-\frac12},\qquad
    \mathsf d\lb z,z'\rb=\sum_{p,q\in\mathbb N+\frac12}\mathsf d_{-q,p}
    z^{-q-\frac12}z'^{-p-\frac12},
  \end{align}
  consider the principal minor expansion of the determinant \eqref{taufredd} and take into account that $\mathsf d_{-q,p}=O\lb 1\rb$ and $\mathsf a_{p,-q}=t^{p+q}\cdot t^{\mathfrak S}O\lb 1\rb t^{-\mathfrak S}$  as $t\to 0$. For instance, under assumption that $|\Re\sigma|<\frac12$ (such a choice is possible for sufficiently generic monodromy) one has
  \beq\label{jimaspv}
  \tau\left[J\right]=1-\operatorname{Tr}\lb\mathsf a_{\frac12,-\frac12}
  \mathsf d_{-\frac12,\frac12}\rb + o\lb t\rb.
  \eeq
  It can be readily shown that
  \ben
  \mathsf a_{\frac12,-\frac12}=t\cdot
  e^{i\pi\mathfrak S}t^{\mathfrak S}
  \lb \begin{array}{cc}
  \frac{\sigma^2+\theta_t^2-\theta_0^2}{2\sigma} & \frac{\lb\theta_t+\sigma\rb^2-\theta_0^2}{2\sigma\lb1-2\sigma\rb}\vspace{0.1cm}\\
  \frac{\theta_0^2-\lb\theta_t-\sigma\rb^2}{2\sigma\lb1+2\sigma\rb}
  & \frac{\theta_0^2-\theta_t^2-\sigma^2}{2\sigma}
  \end{array}\rb
  t^{-\mathfrak S}e^{-i\pi\mathfrak S},\qquad 
  \mathsf d_{-\frac12,\frac12}=s_-^{\sigma_z}\lb \begin{array}{cc}
  -\frac{\theta_*}{2\sigma} & \frac{\sigma-\theta_*}{2\sigma\lb1+2\sigma\rb}\vspace{0.1cm}\\
  \frac{\sigma+\theta_*}{2\sigma\lb1-2\sigma\rb} & \frac{\theta_*}{2\sigma}
  \end{array}\rb s_-^{-\sigma_z},
  \ebn
  so that the asymptotics \eqref{jimaspv} reproduces the result of \cite[Theorem 3.1]{Jimbo}.
  
  In order to obtain complete short-distance expansion of the tau function following the general scheme outlined in \cite[Subsection 2.2]{CGL}\footnote{It generalizes the approach developed earlier in \cite{GL16,GL17} for Painlevé VI and Painlev\'e III ($D_8$).}, one needs to compute all matrix elements $\mathsf a_{p,-q}$, $\mathsf d_{-q,p}$. While this is possible in principle, the result takes a much simpler form if beforehand we make a non-constant scalar gauge transformation of the linear system \eqref{linsys}:
  \beq\label{gtr}
  \Phi\lb z\rb\to \tilde\Phi\lb z\rb = \lb 1-z^{-1}\rb^{-\theta_t}e^{-\frac{tz}{2}}\Phi\lb z\rb.
  \eeq
  The counterpart $\tilde A\lb z,t\rb$ of the matrix $A\lb z,t\rb$ for the transformed system, instead of being traceless, has rank~1 residue at $z=1$ and rank~1 most singular term in the Laurent expansion at $z=\infty$. The transformed Jimbo-Miwa-Ueno tau function can be easily computed along the lines of Subsection~3.1.2, with the result
  \ben
  \tilde{\tau}_{\mathrm{JMU}}\lb t\rb=e^{\theta_t t}{\tau}_{\mathrm{JMU}}\lb t\rb.
  \ebn
  In the gauge theory language, the extra exponential corresponds to the confluent limit of the $\mathrm{U}(1)$-factor. Some readers may also recognize in it the Szeg\H{o} constant in the asymptotics of scalar Toeplitz determinants with the symbol $\lb 1-z^{-1}\rb^{-\theta_t}e^{-\frac{tz}{2}}$.

  The tau functions of the auxiliary systems do not change, whereas the parametrices transform into
  \begin{subequations}\label{tildepsis}
  \begin{align}
  &\begin{aligned}
  &\tilde\Psi_+\lb z\rb = e^{-\frac{tz}{2}}\Psi_+\lb z\rb=\\
    & \quad =e^{i\pi\mathfrak S}t^{\mathfrak S}\lb\begin{array}{cc}
  _1F_1\lb\sigma-\theta_*,2\sigma,-tz\rb & \frac{\lb\theta_*-\sigma\rb tz}{2\sigma\lb 2\sigma+1\rb}\;{}_1F_1\lb1+\sigma-\theta_*,2+2\sigma,-tz\rb \vspace{0.1cm} \\
  \frac{\lb \theta_*+\sigma\rb tz}{2\sigma\lb 2\sigma-1\rb}\;{}_1F_1\lb1-\sigma-\theta_*,2-2\sigma,-tz\rb  & _1F_1\lb-\sigma-\theta_*,-2\sigma,-tz\rb
  \end{array}\rb,
  \end{aligned} \\
  &  \tilde\Psi_-\lb z\rb = \lb 1-z^{-1}\rb^{-\theta_t}\Psi_-\lb z\rb=s_-^{\sigma_z}\lb\begin{array}{cc}
        \phi_{11}\lb z\rb & \phi_{12}\lb z\rb \\
        \phi_{21}\lb z\rb & \phi_{22}\lb z\rb
        \end{array}\rb,
    \end{align}
  \end{subequations}  
 with $\phi_{ij}\lb z\rb$ defined in \eqref{phialbe}. As a consequence, we obtain an alternative Fredholm determinant representation for  $\tau\lb t\rb$, which is less symmetric but more adapted for the derivation of the series representation given in Theorem~B.  
 \begin{lemma}
 We have
 \beq\label{fredd2}
 \tau\lb t\rb=\operatorname{const}\cdot t^{\sigma^2-\theta_0^2-\theta_t^2}e^{-\theta_t t}\operatorname{det}_H\lb\begin{array}{cc}
   \mathbb 1 & \tilde{\mathsf a} \\ \tilde{\mathsf d} & \mathbb 1\end{array}\rb,
 \eeq
 where the operators $\tilde{\mathsf a}$, $\tilde{\mathsf d}$ are obtained from $\mathsf a$, $\mathsf d$ by the replacement of $\Psi_{\pm}\lb z\rb$ in the integral kernels \eqref{adkers} by $\tilde{\Psi}_{\pm}\lb z\rb$ defined  by
 \eqref{tildepsis}.
 \end{lemma}
 
  The transformed parametrices $\tilde{\Psi}_{\pm}\lb z\rb$ satisfy
  \beq\label{difff1}
  \partial_z\tilde{\Psi}_{\pm}\lb z\rb=\tilde{\Psi}_{\pm}\lb z\rb\tilde{A}_{\pm}\lb z\rb-
  z^{-1}\left[\mathfrak S,\tilde{\Psi}_{\pm}\lb z\rb\right],
  \eeq
  where $\tilde{A}_{\pm}\lb z\rb$ are explicitly given by
  \begin{subequations}\label{difff11}
  \begin{align}
  \tilde{A}_{+}\lb z\rb=&\, -\frac{t}{2\sigma}\lb\begin{array}{c}
  \sigma-\theta_* \\
  \sigma+\theta_*
  \end{array}\rb\otimes \lb\begin{array}{cc} 1 & 1\end{array}\rb,\\ 
  \tilde{A}_{-}\lb z\rb=&\,\frac{1}{2\sigma z\lb z-1\rb}\lb\begin{array}{c}
    1\\
    -1
    \end{array}\rb\otimes \lb\begin{array}{cc} \lb\theta_t-\sigma\rb^2-\theta_0^2 & 
    \lb\theta_t+\sigma\rb^2-\theta_0^2\end{array}\rb.
  \end{align}
  \end{subequations}
  The differentiation formulas \eqref{difff1} imply, for instance, that
  \beq\label{difff2}
  \begin{aligned}
  &\lb z\partial_z+z'\partial_{z'}+1\rb \tilde{\mathsf a}\lb z,z'\rb+\left[\mathfrak S,\tilde{\mathsf a}\lb z,z'\rb\right]=-\tilde{\Psi}_{+}\lb z\rb \frac{z\tilde{A}_{+}\lb z\rb-z'\tilde{A}_{+}\lb z'\rb}{z-z'}\tilde{\Psi}_{+}^{-1}\lb z'\rb=\\
  &= \frac{t}{2\sigma}\tilde{\Psi}_{+}\lb z\rb\lb\begin{array}{c}
    \sigma-\theta_* \\
    \sigma+\theta_*
    \end{array}\rb\otimes \lb\begin{array}{cc} 1 & 1\end{array}\rb\tilde{\Psi}_{+}^{-1}\lb z'\rb
    =\frac{t}{2\sigma}\cdot e^{i\pi\mathfrak S}t^{\mathfrak S}u_+\lb z\rb\otimes v_+^T\lb z'\rb t^{-\mathfrak S}e^{-i\pi\mathfrak S},
  \end{aligned}
  \eeq
  with
  \beq\label{difff3}
  u_+\lb z\rb=\lb\begin{array}{c}
  \lb \sigma-\theta_*\rb\;{}_1F_1\lb 1+\sigma-\theta_*,1+2\sigma,-tz\rb \\
  \lb \sigma+\theta_*\rb\;{}_1F_1\lb 1-\sigma-\theta_*,1-2\sigma,-tz\rb
  \end{array}\rb, \qquad  v_+\lb z\rb=\lb\begin{array}{c}
   {}_1F_1\lb 1-\sigma+\theta_*,1-2\sigma,tz\rb \\
   {}_1F_1\lb 1+\sigma+\theta_*,1+2\sigma,tz\rb
    \end{array}\rb.
  \eeq
  Substituting  into \eqref{difff2} the integral kernel $\tilde{\mathsf a}\lb z,z'\rb$ written
  in the Fourier basis and series representations of the hypergeometric functions which appear in \eqref{difff3}, the matrix elements $\tilde{\mathsf a}_{p,-q}$ can be computed in explicit form. Analogous calculation may also be done for $\tilde{\mathsf d}_{-q,p}$. The crucial point here is that the matrices $\tilde{A}_{\pm}\lb z\rb$ in \eqref{difff11} have rank 1; this is why we made the transformation \eqref{gtr}.
 
  Here is the result of the latter calculation:
  \begin{lemma}
  Matrix elements of $\tilde{\mathsf a}$ and $\tilde{\mathsf d}$ in the Fourier basis are given by
  \begin{subequations}
  \begin{align}
  &\begin{aligned}
  \tilde{\mathsf a}_{p,-q}=&\,\frac{\lb-1\rb^{p-\frac12} t^{p+q}}{\lb p-\tfrac12\rb! \lb q-\tfrac12\rb!}\; 
  \operatorname{diag}\left\{
  \tfrac{\lb \sigma-\theta_*\rb_{p+\frac12}e^{i\pi\sigma}t^{\sigma}}{ \lb2\sigma\rb_{p+\frac12}}, 
  \tfrac{\lb -\sigma-\theta_*\rb_{p+\frac12}e^{-i\pi\sigma}t^{-\sigma}}{ \lb -2\sigma\rb_{p+\frac12}}\right\}\times \\
  \times&\,\lb\begin{array}{cc}
  \frac{1}{p+q} & \frac{1}{p+q+2\sigma} \vspace{0.1cm}\\ 
  \frac{1}{p+q-2\sigma} &  \frac{1}{p+q}
  \end{array}\rb
  \operatorname{diag}\left\{\tfrac{\lb 1-\sigma+\theta_*\rb_{q-\frac12}e^{-i\pi\sigma}t^{-\sigma}}{\lb1-2\sigma\rb_{q-\frac12}},
  \tfrac{\lb 1+\sigma+\theta_*\rb_{q-\frac12}e^{i\pi\sigma}t^{\sigma}}{\lb1+2\sigma\rb_{q-\frac12}}\right\},
  \end{aligned}\\
    &\begin{aligned}
    \tilde{\mathsf d}_{-q,p}=&\,\frac{1 }{\lb p-\tfrac12\rb! \lb q-\tfrac12\rb!}\; 
    \operatorname{diag}\left\{
    \tfrac{\lb 1+\theta_0+\theta_t-\sigma\rb_{q-\frac12}
    \lb 1-\theta_0+\theta_t-\sigma\rb_{q-\frac12}s_-}{ \lb1-2\sigma\rb_{q-\frac12}}, 
    -\tfrac{\lb 1+\theta_0+\theta_t+\sigma\rb_{q-\frac12}
    \lb 1-\theta_0+\theta_t+\sigma\rb_{q-\frac12}s_-^{-1}}{ \lb1+2\sigma\rb_{q-\frac12}}\right\}\times \\
    \times&\,\lb\begin{array}{cc}
    \frac{1}{p+q} & \frac{1}{p+q-2\sigma} \vspace{0.1cm}\\ 
    \frac{1}{p+q+2\sigma} &  \frac{1}{p+q}
    \end{array}\rb
    \operatorname{diag}\left\{\tfrac{\lb \theta_0-\theta_t+\sigma\rb_{p+\frac12}\lb -\theta_0-\theta_t+\sigma\rb_{p+\frac12}s_-^{-1}}{\lb2\sigma\rb_{p+\frac12}},
     -\tfrac{\lb \theta_0-\theta_t-\sigma\rb_{p+\frac12}\lb -\theta_0-\theta_t-\sigma\rb_{p+\frac12}s_-}{\lb -2\sigma\rb_{p+\frac12}}\right\},
    \end{aligned}
  \end{align}
  \end{subequations}
    where  $\lb\alpha\rb_n=\frac{\Gamma\lb \alpha +n\rb}{\Gamma\lb a\rb}$ is the Pochhammer symbol. 
  \end{lemma}
  
  Since $\tilde{\mathsf a}$ and $\tilde{\mathsf d}$ are represented, up to diagonal factors, by (semi-infinite) Cauchy matrices, all of their minors can be computed in a factorized form. To finish the proof of Theorem~B, it now suffices to match the individual terms in the principal minor expansion of the determinant \eqref{fredd2} with those of the Fourier transformed combinatorial expansion \eqref{confCB1series} with $c=1$. In this context, half-integer momenta such as $p$ and $q$ play the role of Frobenius-type coordinates of charged Young diagrams; the reader is referred to \cite[Sub\-section~2.2]{CGL} and \cite[Appendix A]{GL16} for the details.
  
  \subsection{Connection formulae between $0$ and $i\infty$ from confluent CBs}
  Our starting point is the $s$-channel representation of the Painlev\'e VI tau function \cite{GIL12}:
    \beq\label{schexp}
    \tau_{\mathrm{VI}}\left(t\right)\equiv \tau_{\mathrm{VI}}\lb\substack{\theta_{1}\;\qquad\; 
           \theta_{t}\\ \sigma_{0t},s_{0t} \\ \theta_{\infty}\qquad\; \theta_0};t\rb =\sum_{n\in\mathbb Z}
    \mathcal C_{\mathrm{VI}}\lb\substack{\theta_{1}\;\qquad\; 
           \theta_{t}\\ \sigma_{0t}+n \\ \theta_{\infty}\qquad\; \theta_0}\rb s_{0t}^n\,
    \mathcal F\lb\substack{\theta_{1}\;\qquad\; 
       \theta_{t}\\ \sigma_{0t}+n \\ \theta_{\infty}\qquad\; \theta_0};t\rb.
    \eeq
  Here $\mathcal F\lb t\rb$ is the regular Virasoro conformal block with $c=1$. The latter condition on the central charge is assumed throughout the whole subsection. The structure constants are expressed in terms of the Barnes $G$-functions:
  \ben
      \mathcal C_{\mathrm{VI}}\lb\substack{\theta_{1}\;\quad 
             \theta_{t}\\ \sigma \\ \theta_{\infty}\quad \theta_0}\rb=
       \frac{\prod_{\epsilon,\epsilon'=\pm}G\left(
       1+\theta_t+\epsilon\theta_0+\epsilon'\sigma\right)
       G\left(
           1+\theta_1+\epsilon\theta_{\infty}+\epsilon'\sigma\right)}{
           \prod_{\epsilon=\pm}G\left(1+2\epsilon\sigma\right)}. 
           \ebn
 The parameters $\lb\sigma_{0t},s_{0t}\rb$ are Painlevé VI integrals of motion which can be explicitly related to monodromy data of the associated linear system. The rigorous proof of \eqref{schexp}, where $\mathcal F\lb t\rb$ is interpreted as combinatorial series \eqref{CBseries}, was given in \cite{GL16};  our Theorem~B is a PV analog of this result. In the present subsection, we are going to adopt in addition the usual CFT assumptions on the analytic properties of CBs and consider $\mathcal F\lb t\rb$ as a function that can be continued
  from the cut disc $|t|<1$, $t\notin(-1,0]$ to the cut plane ${t\in\Cb\backslash\lb\lb -\infty,0]\cup[1,\infty\rb\rb }$.
  
  The crucial fact is the existence of the confluent limit \eqref{confCB1series}. It allows us to recover the short-distance expansion \eqref{fourier0} of the PV tau function as
  \beq\label{conflim}
        \tau\lb t\rb=\lim_{\Lambda\to\infty}\;\frac{\Lambda^{-\theta_0^2-\theta_t^2}}{
         G^2\left(1+\Lambda\right)}\tau_{\mathrm{ VI}}\left(\frac t\Lambda\right),
   \eeq      
   under identification $\theta_1=\frac{\Lambda+\theta_*}{2}$, $\theta_{\infty}=\frac{\Lambda-\theta_*}{2}$, $\sigma=\sigma_{0t}$, $e^{2\pi i \eta}=s_{0t}$ and normalization $\mathcal N_0=1$. Indeed, the large argument asymptotics of the Barnes function further implies that, as $\Lambda\to\infty$,
     \beq\label{barnesasy}
     \prod_{\epsilon=\pm}\frac{G\left(1+\Lambda+\epsilon\alpha\right)}{
     G\left(1+\Lambda+\epsilon\beta\right)}=\Lambda^{\alpha^2-\beta^2}\left[1+O\left(\Lambda^{-2}\right)\right].
     \eeq
     The latter result is valid in any sector of the complex  $\Lambda$-plane not containing the negative real axis, i.e. the fractional power corresponds to $\arg\Lambda\in\lb-\pi,\pi\rb$. The asymptotics \eqref{barnesasy} immediately gives
     \ben
     \mathcal C_0\lb \theta_*;\sigma;\substack{\theta_t \\ \theta_0}\rb=
     \lim_{\Lambda\to\infty}\frac{\Lambda^{-\sigma^2}}{G^2\left(1+\Lambda\right)} \mathcal C_{\mathrm{VI}}\lb\substack{\frac{\Lambda+\theta_*}{2}\quad 
                   \theta_{t}\\ \quad\sigma \\ \frac{\Lambda-\theta_*}{2}\quad \theta_0}\rb.
     \ebn  
     
     On the other hand, one could write an expansion similar to \eqref{schexp} with $t$ replaced by any of the remaining 5~cross-ratios $\frac{t}{t-1}$, $1-t$, $\frac{t-1}{t}$, $\frac1{1-t}$, $\frac1t$ after suitable transformation of parameters. These transformations of the tau function $\tau_{\mathrm{VI}}\lb t\rb$ are generated by two basic ones: the braiding and fusion relation. 
     
     \begin{lemma}
     We have 
     \beq\label{braidingPVItau}
     \tau_{\mathrm{VI}}\lb\substack{\theta_{1}\;\qquad\; 
                \theta_{t}\\ \sigma_{0t},s_{0t} \\ \theta_{\infty}\qquad\; \theta_0};t\rb=
     \chi_{\mathrm{b}}^{\pm}\cdot \lb1-t\rb^{-2\theta_t^2}
     \tau_{\mathrm{VI}}\lb\substack{\theta_{\infty}\qquad 
                 \theta_{t}\\ \;\;\sigma_{0t},\tilde s_{0t}^{\pm}\;\; \\ \theta_{1}\qquad\;\;\; \theta_0};\frac{t}{t-1}\rb,  \qquad \Im t\gtrless 0,               
     \eeq
     where 
     \begin{subequations}
     \begin{align}
     \tilde s_{0t}^{\pm}= &\,\frac{\sin\pi \lb\theta_{\infty}-\theta_1+\sigma_{0t}\rb}{\sin\pi \lb\theta_{\infty}-\theta_1-\sigma_{0t}\rb}\,e^{\pm 2\pi i \sigma_{0t}}s_{0t},\\
     \chi_{\mathrm{b}}^{\pm}\equiv &\,\chi_{\mathrm{b}}^{\pm}\lb\substack{\theta_{1}\;\quad 
     \theta_{t}\\ \sigma_{0t} \\ \theta_{\infty}\quad \theta_0}\rb=\frac{\hat{G}\lb\theta_1-\theta_{\infty}+\sigma_{0t}\rb}{
     \hat{G}\lb\theta_{\infty}-\theta_{1}+\sigma_{0t}\rb}\,
     e^{\pm i\pi\lb\sigma_{0t}^2-\theta_0^2-\theta_t^2\rb},
     \end{align}
     \end{subequations}
     and the fractional powers are given by their principal branches.
     \end{lemma}
     \pf Use the braiding transformation of regular CBs and the relation
     \ben
   \left.  \mathcal C_{\mathrm{VI}}\lb\substack{\theta_{1}\;\qquad\; 
                \theta_{t}\\ \sigma_{0t}+n \\ \theta_{\infty}\qquad\; \theta_0}\rb\right/
     \mathcal C_{\mathrm{VI}}\lb\substack{\theta_{\infty}\qquad 
                     \theta_{t}\\ \sigma_{0t}+n \\ \theta_{1}\qquad\; \theta_0} \rb 
     =\frac{\hat{G}\lb\theta_1-\theta_{\infty}+\sigma_{0t}+n\rb}{
     \hat{G}\lb\theta_{\infty}-\theta_1+\sigma_{0t}+n\rb}  =\frac{\hat{G}\lb\theta_1-\theta_{\infty}+\sigma_{0t}\rb}{
          \hat{G}\lb\theta_{\infty}-\theta_1+\sigma_{0t}\rb} \lb\frac{\sin \pi\lb\theta_{\infty}-\theta_1+\sigma_{0t}\rb }{\sin\pi\lb\theta_1-\theta_{\infty}+\sigma_{0t}\rb}\rb^n,                       
     \ebn
     where $\hat{G}\lb z\rb=\ds\frac{G\lb 1+z\rb}{G\lb 1-z\rb}$ and the last equality follows from the 
     recurrence relation $\hat{G}\lb z+1\rb=-\ds\frac{\pi}{\sin\pi z}\hat{G}\lb z\rb$.\epf
     
     The fusion relation  was first obtained in \cite{ILT13}. To write it explicitly, we need to introduce some further notation. Define the following functions of Painlevé VI parameters:
     \beq
     \begin{gathered}
     p_{\nu}:=2\cos2\pi\theta_{\nu},\qquad \nu=0,t,1,\infty,\\
     \omega_{0t}:=p_0p_t+p_1p_{\infty},\qquad \omega_{1t}:=p_1p_t+p_0p_{\infty},\qquad
     \omega_{01}:=p_0p_1+p_tp_{\infty}.
     \end{gathered}
     \eeq
     It can be easily checked that $\omega_{0t}$, $\omega_{1t}$, $\omega_{01}$ are invariant under the $W\lb D_4\rb$ transformations described in the introduction. For given $\sigma_{0t}$, $s_{0t}$, let 
     $\sigma_{1t}$, $\sigma_{01}$ be (arbitrarily chosen) solutions of the equations 
     \ben
     \begin{gathered}
     4\sin^22\pi\sigma_{0t}\cos2\pi \sigma_{1t}=\omega_{1t}-\omega_{01}\cos2\pi\sigma_{0t}-
     8\sum_{\epsilon=\pm1}s_{0t}^{\epsilon}\prod_{\epsilon'=\pm1}\sin\pi\lb\theta_1+\epsilon'\theta_{\infty}-\epsilon\sigma_{0t}\rb \sin\pi\lb\theta_t+\epsilon'\theta_{0}-\epsilon\sigma_{0t}\rb,\\
     4\sin^22\pi\sigma_{0t}\cos2\pi \sigma_{01}=\omega_{01}-\omega_{1t}\cos2\pi\sigma_{0t}+
     8\sum_{\epsilon=\pm1}s_{0t}^{\epsilon}e^{-2\pi i \epsilon\sigma_{0t}}\!\!\!\prod_{\epsilon'=\pm1}\sin\pi\lb\theta_1+\epsilon'\theta_{\infty}-\epsilon\sigma_{0t}\rb \sin\pi\lb\theta_t+\epsilon'\theta_{0}-\epsilon\sigma_{0t}\rb.
     \end{gathered}    
     \ebn
     As the notation suggests, these quantities represent the exponents of composite monodromy. Also, denote    
     \ben
     s_{1t}:=\frac{\lb4\cos2\pi\sigma_{01}+4\cos2\pi\sigma_{0t}\cos2\pi\sigma_{1t}-\omega_{01}\rb
     e^{-2\pi i \sigma_{1t}}+
     \lb4\cos2\pi\sigma_{0t}+4\cos2\pi\sigma_{01}\cos2\pi\sigma_{1t}-\omega_{0t}\rb}{
     16\prod\limits_{\epsilon=\pm1}\sin\pi\lb\theta_t-\sigma_{1t}+\epsilon\theta_1\rb
     \sin\pi\lb\theta_0-\sigma_{1t}+\epsilon\theta_{\infty}\rb  }.
     \ebn
     \begin{lemma}[{\cite[Theorem A]{ILP}}]
     We have
     \beq\label{fusionPVItau}
     \tau_{\mathrm{VI}}\lb\substack{\theta_{1}\;\qquad\; 
                \theta_{t}\\ \sigma_{0t},s_{0t} \\ \theta_{\infty}\qquad\; \theta_0};t\rb=
     \chi_{\mathrm{f}}\cdot
     \tau_{\mathrm{VI}}\lb\substack{\theta_{0}\qquad \;
                 \theta_{t}\\ \;\;\sigma_{1t},s_{1t}\; \\ \theta_{\infty}\qquad\; \theta_1};1-t\rb,               
     \eeq     
     where the connection constant $\chi_{\mathrm{f}}$ is given by
     \beq
     \begin{aligned}
     \chi_{\mathrm{f}}\equiv \chi_{\mathrm{f}}\lb \substack{\theta_{1}\;\qquad\; 
                     \theta_{t}\\ \sigma_{0t},s_{0t} \\ \theta_{\infty}\qquad\; \theta_0}\rb=&\,
                     \prod_{\epsilon,\epsilon'=\pm1}
                           \frac{G\left( 1+\epsilon\sigma_{1t}+\epsilon'\theta_t-
                               \epsilon\epsilon'\theta_1\right) G\left(1+\epsilon\sigma_{1t}+\epsilon'\theta_0-
                                   \epsilon\epsilon' \theta_{\infty}\right)}{
                           G\left(1+\epsilon\sigma_{0t}+\epsilon'\theta_t+\epsilon\epsilon'\theta_0\right)   G\left(1+\epsilon\sigma_{0t}+\epsilon'\theta_1+\epsilon\epsilon'
                               \theta_{\infty}\right)}\;\times\\
                               &\; \times
                          \prod_{\epsilon=\pm1}
                          \frac{G(1+2\epsilon\sigma_{0t})}{G(1+2\epsilon\sigma_{1t})}
                          \prod_{k=1}^4\frac{\hat{G}(\varsigma +\nu_k)}{ \hat{G}(\varsigma +\lambda_k)}.
                          \end{aligned}
     \eeq
      The parameters $\nu_{1},\ldots ,\nu_4$ and $\lambda_{1},\ldots,\lambda_4$ are defined by
     \beq
       \begin{aligned}\label{nus}
         \begin{array}{ll}
         \nu_1=\sigma_{0t}+\theta_0+\theta_t, & \qquad\lambda_1=\theta_0+\theta_t+\theta_1+\theta_{\infty},\\
         \nu_2=\sigma_{0t}+\theta_1+\theta_{\infty},& \qquad\lambda_2=\sigma_{0t}+\sigma_{1t}+\theta_0+\theta_1, \\
         \nu_3=\sigma_{1t}+\theta_0+\theta_{\infty},& \qquad\lambda_3=\sigma_{0t}+\sigma_{1t}+\theta_t+\theta_{\infty},\\
         \nu_4=\sigma_{1t}+\theta_t+\theta_1,& \qquad\lambda_4=0,\\
      \end{array}
         \end{aligned}
      \eeq
      and the quantity $\varsigma$ is determined by
      \beq\label{eqvarsigma}
      e^{2\pi i \varsigma}=\frac{2\cos2\pi\lb \sigma_{0t}-\sigma_{1t}\rb 
       -2\cos2\pi\lb\theta_0+\theta_1\rb
       -2\cos2\pi\lb\theta_{\infty}+\theta_t\rb
        +2\cos2\pi\sigma_{01}}{\sum_{k=1}^4\lb e^{2\pi i \lb\nu_{\Sigma}-\nu_k\rb}
        - e^{2\pi i \lb\nu_{\Sigma}-\lambda_k\rb}\rb},
      \eeq
      with $2\nu_{\Sigma}=\sum_{k=1}^4\nu_k=\sum_{k=1}^4\lambda_k$.
     \end{lemma}
 
    Combining two braiding and one fusion transformation as 
    $t\to \frac{t}{t-1}\to \frac{1}{1-t}\to \frac1t$, one can relate the $s$- and $u$-channel expansions of the PVI tau function.
    \begin{cor}
    One can write, e.g.
    \beq\label{sutau}
    \tau_{\mathrm{VI}}\lb\substack{\theta_{1}\;\qquad\; 
                    \theta_{t}\\ \sigma_{0t},s_{0t} \\ \theta_{\infty}\qquad\; \theta_0};t\rb=
    \chi_{0\infty}\cdot t^{-2\theta_t^2}\tau_{\mathrm{VI}}\lb\substack{\theta_{1}\;\qquad\; 
                     \theta_{t}\\ \;\tilde\sigma_{01},\tilde s_{01}\;\; \\ \theta_{0}\qquad\;\;\; \theta_{\infty}};\frac1t\rb,       \qquad \Im t> 0,           
    \eeq
    where $\tilde\sigma_{01}$ is an arbitrary solution of
  \begin{align}\label{sigma01ref}
  \begin{aligned}
  4\sin^22\pi\sigma_{0t}\cos2\pi\tilde\sigma_{01}=&\,\omega_{01}-
  \omega_{1t}\cos2\pi\sigma_{0t}+\\
  +&\,8\sum_{\epsilon=\pm}
  s_{0t}^{\epsilon}e^{2\pi i
  \epsilon\sigma_{0t}}\prod_{\epsilon'=\pm}
  \sin\pi\left(\theta_1-\epsilon\sigma_{0t}+\epsilon'\theta_{\infty}
  \right)
  \sin\pi\left(\theta_t-
  \epsilon\sigma_{0t}+\epsilon'\theta_0\right),
  \end{aligned}
  \end{align}
  and the parameter $\tilde{s}_{01}$ (for chosen $\tilde{\sigma}_{01}$) is determined by
   \begin{align}\label{s01ref}
   \tilde s_{01}=\frac{e^{2\pi i \tilde\sigma_{01}}
   \left( Q_+ s_{0t}+  Q_- s_{0t}^{-1}+  Q_0\right)}{8\sin^4 2\pi \sigma_{0t}\prod_{\epsilon=\pm}
    \sin\pi\left(\theta_1-\tilde\sigma_{01}+\epsilon\theta_0
    \right)
    \sin\pi\left(\theta_t-
    \tilde\sigma_{01}+\epsilon\theta_{\infty}\right)},
   \end{align}
   with
   \begin{subequations}
   \begin{align}\label{coefsQ}
   &\begin{aligned}
   Q_{\pm}=& \;e^{\pm 2\pi i \sigma_{0t}}\left(\omega_{1t}-\omega_{01}\cos2\pi\sigma_{0t}-
   4\sin^22\pi\sigma_{0t}\cos2\pi\left(\tilde\sigma_{01}\mp\sigma_{0t}\right)
   \right)\times\\
   \times&\;\prod\nolimits_{\epsilon=\pm}
   \sin\pi\left(\theta_t\mp\sigma_{0t}
     +\epsilon\theta_{0}\right)
   \sin\pi\left(\theta_1\mp\sigma_{0t}
   +\epsilon\theta_{\infty}\right),
   \end{aligned}
   \\
   &8Q_0=\omega_{01}\omega_{1t}\left(1+\cos^2 2\pi\sigma_{0t}\right)-\left(\omega_{01}^2+\omega_{1t}^2\right)
      \cos2\pi\sigma_{0t}+4\sin^42\pi\sigma_{0t}\lb4\cos2\pi\sigma_{0t}-\omega_{0t}\rb.
   \end{align}
   \end{subequations}
  The connection coefficient $\chi_{0\infty}$ is expressed as
  \ben
  \chi_{0\infty}=\chi_{\mathrm{b}}^{+}\lb\substack{\theta_{1}\;\quad 
       \theta_{t}\\ \sigma_{0t} \\ \theta_{\infty}\quad \theta_0}\rb
  \chi_{\mathrm{f}}\lb \substack{\theta_{\infty}\qquad 
                       \theta_{t}\\ \sigma_{0t},\tilde s_{0t}^+ \\ \;\theta_{1}\qquad \;\;\theta_0}\rb     
  \chi_{\mathrm{b}}^{+}  \lb\substack{\theta_{0}\;\quad\; 
                       \theta_{t}\\ \;\tilde\sigma_{01}\; \\ \;\theta_{1}\quad\; \theta_{\infty}}\rb.   
  \ebn
    \end{cor}   
  \begin{rmk}
   It should be stressed that $\tilde{\sigma}_{01}\ne\sigma_{01}$. In fact, \ben
    2\cos2\pi\sigma_{01}+2\cos2\pi\tilde\sigma_{01}=\omega_{01}-
    4\cos2\pi\sigma_{0t}\cos2\pi\sigma_{1t}.\ebn
  \end{rmk}    
  \begin{rmk} Let us emphasize that the CB series appearing in the two sides of \eqref{sutau} have complementary domains of convergence, given by the interior and exterior of the unit circle $|t|=1$. The tau functions $\tau_{\mathrm{VI}}\lb t\rb$ and $\tau_{\mathrm{VI}}\lb t^{-1}\rb$ should therefore be interpreted as analytic continuations from these domains which can be compared on $\mathbb C\backslash\mathbb R$.
  Having in mind further application to Painlevé V, we restricted our attention to $\Im t>0$, but similar formulas can also be derived for lower half-plane.

  \end{rmk}  
  We are now in the position to analyze the confluent limit of the relation \eqref{sutau}. We have already identified above the procedure which gives on the left the short-distance expansion of the PV tau function $\tau\lb t\rb$ with prescribed monodromy data $\lb \sigma,\eta\rb$. The corresponding limit of the right side of \eqref{sutau} yields
  \beq\label{stranlim}
  \lim_{\Lambda\to\infty}\;\frac{\Lambda^{-\theta_0^2-\theta_t^2} }{
           G^2\left(1+\Lambda\right)}\;\chi_{0\infty}\, \lb\frac{\Lambda}{t}\rb^{2\theta_t^2}
  \sum_{n\in\mathbb Z} 
  \mathcal{C}_{\mathrm{VI}}\lb\substack{\frac{\Lambda+\theta_*}{2}\;\;\qquad\quad 
   \theta_{t}\;\;\\  \tilde\sigma_{01}+n \\ 
  \;\;\;\theta_0\qquad\quad \frac{\Lambda-\theta_*}{2}}\rb\,
  \tilde{s}_{01}^n\,\mathcal F\lb\substack{\frac{\Lambda+\theta_*}{2}\;\;\qquad\quad 
                          \theta_{t}\;\;\\  \tilde\sigma_{01}+n \\ 
                          \;\;\;\theta_0\qquad\quad \frac{\Lambda-\theta_*}{2}};\frac{\Lambda}{t}\rb,
  \eeq
  where, as before, $\theta_1=\frac{\Lambda+\theta_*}{2}$, $\theta_{\infty}=\frac{\Lambda-\theta_*}{2}$ and $\sigma_{0t}=\sigma$, 
  $s_{0t}=e^{2\pi i\eta}$ remain finite. 
  
  It will become clear in a moment that, for efficient computation of the asymptotics of the structure constants using e.g. \eqref{barnesasy}, one needs to assume that $\arg\Lambda\ne 0,\pm\pi$. Below will need to interpret expressions such as $\lb1-\frac{\Lambda}{t}\rb^{\alpha}$. Since our ultimate goal is to derive from \eqref{stranlim} the asymptotic expansion of $\tau\lb t\rb$ as
  $t\to i\infty$, the limit $\Lambda\to \infty$ should be supplemented with an extra condition $-\pi<\arg\Lambda<0$ which will be always understood. It follows e.g. that $2\cos\pi\lb\Lambda+\alpha\rb\sim e^{i\pi\lb\Lambda+\alpha\rb}$, so that
  \beq
  \omega_{1t}\sim e^{i\pi\Lambda}A_-,\qquad   
  \omega_{01}\sim e^{i\pi\Lambda}A_+,\qquad \tilde{\sigma}_{01}\sim\frac{\Lambda}{2}+\nu,
  \eeq
  where $A_{\pm}$ are given by \eqref{Apm} and $\nu$ by \eqref{nudefi}, \eqref{zetap}. This in turn allows us to find the asymptotics of 
  $\tilde{s}_{01}$: 
   \begin{align}
   \tilde{s}_{01}&\sim-e^{2\pi i \Lambda}\frac{i e^{i\pi\left(\nu+\theta_t+\frac{\theta_*}{2}\right)}
   \left( R_+ e^{2\pi i \eta}+  R_- e^{-2\pi i \eta}+  R_0\right)}{
   8\sin^42\pi\sigma
   \sin\pi\left(\theta_t-\nu-\frac{\theta_*}{2}\right)
   \prod_{\epsilon=\pm}\sin\pi\left(\frac{\theta_*}{2}-\nu+\epsilon\theta_0\right)}, 
   \end{align}
   where the coefficients $R_{\pm,0}$ are given by
   \begin{align*}
   R_{\pm}&=\lb A_+-A_-\cos2\pi\sigma
    -2\sin^22\pi\sigma e^{2\pi i\left(\nu\mp \sigma\right)}\rb \left(-i\right)e^{\pm i\pi\sigma}\sin\pi\left(\theta_*\mp\sigma\right)
   \prod_{\epsilon=\pm}
    \sin\pi\left(\theta_t\mp\sigma
      +\epsilon\theta_{0}\right) ,\\
   R_0&=\left(\cos^22\pi\theta_0+\cos^22\pi\theta_t\right)
   \left(1+\cos^22\pi\sigma-2\cos2\pi\sigma\cos2\pi\theta_*\right)-
   \sin^42\pi\sigma+\\   
   &+2\cos2\pi\theta_0\cos2\pi\theta_t\left[\cos2\pi\theta_*\left(
   1+\cos^22\pi\sigma\right)-2\cos2\pi\sigma\right].
   \end{align*}
   One can also show by straightforward but rather tedious algebra that 
   \beq
   -\frac{R_+ e^{2\pi i \eta}+  R_- e^{-2\pi i \eta}+  R_0}{\sin^42\pi\sigma}=e^{2\pi i \rho},
   \eeq
   where $\rho$ is also defined by \eqref{nudefi}, \eqref{zetap}.
   
   Our next task is to analyze the asymptotics of structure constants. From the estimate $\tilde{\sigma}_{01}\sim\frac{\Lambda}{2}+\nu$  it may be further deduced that
   \begin{align*}
   \mathcal{C}_{\mathrm{VI}}\lb\substack{\frac{\Lambda+\theta_*}{2}\;\;\qquad\quad 
      \theta_{t}\;\;\\  \tilde\sigma_{01}+n \\ 
     \;\;\;\theta_0\qquad\quad \frac{\Lambda-\theta_*}{2}}\rb\sim&\;
     \lb 2\pi\rb^{2\lb\nu+n\rb}\mathcal C_{i\infty}\lb\substack{\theta_t\\ \theta_*};\nu+n;\theta_0\rb
     \left[\hat G\lb \nu+n-\theta_t+\tfrac{\theta_*}{2}\rb\prod_{\epsilon=\pm}
     \hat G\lb \nu+n+\epsilon\theta_0-\tfrac{\theta_*}{2}\rb\right]^{-1}
     \times\\
     \times&\;\frac{\prod_{\epsilon=\pm}G\lb 1+\Lambda+\nu+n+\epsilon\theta_t-\frac{\theta_*}{2}\rb
     G\lb 1+\Lambda+\nu+n+\epsilon\theta_0+\frac{\theta_*}{2}\rb}{G^2\lb1+\Lambda\rb G^2\lb1+\Lambda+2\nu+2n\rb}\times\\
     \times&\; G^2\lb1+\Lambda\rb \frac{\hat{G}\lb\Lambda+2\nu+2n\rb}{
     \hat{G}\lb\Lambda+\nu+n-\theta_t-\frac{\theta_*}{2}\rb},
   \end{align*}
   where $\mathcal C_{i\infty}$ is defined by \eqref{strfourier8}. The first line of the above relation is finite. The second can be shown to be asymptotic to $\Lambda^{\theta_0^2+\theta_t^2+\frac{\theta_*^2}{2}-2\lb\nu+n\rb^2}$ using 
   \eqref{barnesasy}. The ratio of two $\hat{G}$-fuctions in the third line can be estimated using that for $\arg \Lambda\in\lb -\pi,0\rb$ we have $\ds\frac{\hat{G}\lb\Lambda+\alpha\rb}{\hat{G}\lb\Lambda+\beta\rb}
   \sim \lb2\pi\rb^{\alpha-\beta} e^{-i\pi\Lambda\lb\alpha-\beta\rb-\frac{i\pi}{2}\lb\alpha^2-\beta^2\rb}$, whereas the remaining factor $G^2\lb 1+\Lambda\rb$ will be absorbed later in the limit \eqref{conflim}. Using in addition the recurrence relation  $\hat{G}\lb z+1\rb=-\ds\frac{\pi}{\sin\pi z}\hat{G}\lb z\rb$, the last relation yields
   \beq\label{asy000}
   \begin{gathered}
   \frac{\Lambda^{-\theta_0^2-\theta_t^2} }{
              G^2\left(1+\Lambda\right)}\; 
     \mathcal{C}_{\mathrm{VI}}\lb\substack{\frac{\Lambda+\theta_*}{2}\;\;\qquad\quad 
      \theta_{t}\;\;\\  \tilde\sigma_{01}+n \\ 
     \;\;\;\theta_0\qquad\quad \frac{\Lambda-\theta_*}{2}}\rb\,
     \tilde{s}_{01}^n\,\mathcal F\lb\substack{\frac{\Lambda+\theta_*}{2}\;\;\qquad\quad 
                             \theta_{t}\;\;\\  \tilde\sigma_{01}+n \\ 
   \;\;\;\theta_0\qquad\quad \frac{\Lambda-\theta_*}{2}};\frac{\Lambda}{t}\rb\sim
   \Lambda^{\frac{\theta_*^2}{2}-2\lb\nu+n\rb^2}
      e^{i\pi\lb n-\nu-\theta_t-\frac{\theta_*}{2}\rb\Lambda}e^{-2\pi i n\nu}     
       \times\\
   \qquad\times     \frac{\lb -1\rb^n
      \lb2\pi\rb^{3\nu+\theta_t+\frac{\theta_*}{2}}
 e^{-\frac{i\pi}{2}\lb\nu+\theta_t+\frac{\theta_*}{2}\rb
    \lb 3\nu-\theta_t-\frac{\theta_*}{2}\rb}     
     }{\hat G\lb \nu-\theta_t+\tfrac{\theta_*}{2}\rb\prod_{\epsilon=\pm}
          \hat G\lb \nu+\epsilon\theta_0-\tfrac{\theta_*}{2}\rb}
     \mathcal C_{i\infty}\lb\substack{\theta_t\\ \theta_*};\nu+n;\theta_0\rb e^{2\pi i n\rho} \mathcal F\lb\substack{\frac{\Lambda+\theta_*}{2}\;\;\qquad\quad 
                                  \theta_{t}\;\;\\  \frac{\Lambda}{2}+\nu+n \\ 
        \;\;\;\theta_0\qquad\quad \frac{\Lambda-\theta_*}{2}};\frac{\Lambda}{t}\rb. 
        \end{gathered}  
   \eeq
   Now recall that, according to \eqref{limirr} with $c=1$, we have
    \begin{align*}
    \lim_{\Lambda\to\infty}\lb\frac t\Lambda\rb^{
    \lb\frac{\theta_*}{2}+\nu+n\rb \Lambda-\frac{\theta_*^2}{4}-\theta_t^2+\lb\nu+n\rb^2}\lb 1-\frac {\Lambda}t\rb^{\lb\frac{\theta_*}{2}+\nu+n\rb \Lambda+\frac{\theta_*^2}{4}+\theta_t^2-\lb\nu+n\rb^2}
     \mathcal F\lb\substack{\frac{\Lambda+\theta_*}{2}\;\;\qquad 
                  \theta_{t}\;\;\\  \frac{\Lambda}{2}+\nu+n \\ 
                  \;\;\;\theta_0\qquad \frac{\Lambda-\theta_*}{2}};\frac{\Lambda}{t}\rb=\hat{\mathcal D}\lb\substack{\theta_t\\ \theta_*};\nu+n;\theta_0;t\rb,
    \end{align*}
    at least in the sense of formal series $\Cb[[t^{-1}]]$. Inserting appropriate factor into \eqref{asy000}, dividing by it back, and further interpreting
  $
   \lb\frac t\Lambda\rb^{-
       \lb\frac{\theta_*}{2}+\nu+n\rb \Lambda+\frac{\theta_*^2}{4}+\theta_t^2-\lb\nu+n\rb^2}\lb 1-\frac {\Lambda}t\rb^{-\lb\frac{\theta_*}{2}+\nu+n\rb \Lambda-\frac{\theta_*^2}{4}-\theta_t^2+\lb\nu+n\rb^2}$
   as
   \begin{align*}
   \lb\frac t\Lambda\rb^{2\theta_t^2+\frac{\theta_*^2}{2}-2\lb\nu+n\rb^2}\lb1-\frac{t}{\Lambda}\rb^{-\lb\frac{\theta_*}{2}+\nu+n\rb \Lambda-\frac{\theta_*^2}{4}-\theta_t^2+\lb\nu+n\rb^2}e^{-i\pi\lb
   \lb\frac{\theta_*}{2}+\nu+n\rb \Lambda+\frac{\theta_*^2}{4}+\theta_t^2-\lb\nu+n\rb^2\rb}\sim\\
   \sim
    \lb\frac t\Lambda\rb^{2\theta_t^2+\frac{\theta_*^2}{2}-2\lb\nu+n\rb^2}e^{-i\pi\lb
       \lb\frac{\theta_*}{2}+\nu+n\rb \Lambda+\frac{\theta_*^2}{4}+\theta_t^2-\lb\nu+n\rb^2\rb}
    e^{\lb \frac{\theta_*}{2}+\nu+n\rb t},
   \end{align*}
   we may rewrite the limit \eqref{stranlim} as
   \begin{align*}
   \frac{
            \lb2\pi\rb^{3\nu+\theta_t+\frac{\theta_*}{2}}
       e^{i\pi\lb\nu^2-\theta_t^2-\frac{\theta_*^2}{2}\rb-\frac{i\pi}{2}\lb\nu+\theta_t+\frac{\theta_*}{2}\rb
          \lb 3\nu-\theta_t-\frac{\theta_*}{2}\rb}     
           }{\hat G\lb \nu-\theta_t+\tfrac{\theta_*}{2}\rb\prod_{\epsilon=\pm}
                \hat G\lb \nu+\epsilon\theta_0-\tfrac{\theta_*}{2}\rb}
   \lim_{\Lambda\to\infty}
   e^{-i\pi\Lambda\lb2\nu+\theta_t+\theta_*\rb}\chi_{0\infty}\;
   \times \\ \times\sum_{n\in Z}    \mathcal C_{i\infty}\lb\substack{\theta_t\\ \theta_*};\nu+n;\theta_0\rb e^{2\pi i n\rho}t^{\frac{\theta_*^2}{2}-2\lb\nu+n\rb^2}e^{\lb \frac{\theta_*}{2}+\nu+n\rb t}\hat{\mathcal D}\lb\substack{\theta_t\\ \theta_*};\nu+n;\theta_0;t\rb.
   \end{align*}
   We have thus reproduced the Fourier expansion \eqref{fourier8} and obtained the connection formulas \eqref{nudefi} of Conjecture C. Computing the limit in the first line (it is rather involved but nevertheless straightforward), we also obtain the connection constant $ \Upsilon_{0\to i\infty}$ given in the equation \eqref{Ups0i8} of Conjecture~D.
   
   The above derivation of the connection formulae is admittedly formal. However, we have checked the final results by integrating Painlevé V equation numerically for randomly chosen parameter values and found perfect agreement.

  \subsection{More on Conjecture D}
  We are going to obtain the second connection constant \eqref{Upsi88} using the method of \cite{ILT13,ILT14}. Consider $ \Upsilon_{i\infty\to +\infty}$ as a function of monodromy parameters $\nu,\omega$ defined in \eqref{nup}, \eqref{nudefi}. The Fourier transform structure of the asymptotic expansions \eqref{fourier8}, \eqref{fourier8r} on the imaginary and real axis implies the recurrence relations
  \begin{subequations}
  \begin{align}
  \frac{\Upsilon_{i\infty\to +\infty}\lb\nu+1,\omega\rb}{\Upsilon_{i\infty\to +\infty}\lb\nu,\omega\rb}=&\,e^{-2\pi i\rho}=e^{-2\pi i\omega},\\
  \frac{\Upsilon_{i\infty\to +\infty}\lb\nu,\omega+1\rb}{\Upsilon_{i\infty\to +\infty}\lb\nu,\omega\rb}=&\; e^{2\pi i \xi}=e^{-2\pi i \nu}\lb 1-e^{2\pi i\omega}\rb.
  \end{align}
  \end{subequations}
  The general solution of these equations can be written as
  \beq
  \Upsilon_{i\infty\to +\infty}\lb\nu,\omega\rb = \lb 2\pi \rb^{\omega}e^{\frac{\pi i \omega^2}{2}-2\pi i\omega\nu}\hat{G}\lb -\omega\rb \Upsilon_{\mathrm{per}}\lb\nu,\omega\rb,
  \eeq
  where $\Upsilon_{\mathrm{per}}\lb\nu,\omega\rb$ is an arbitrary function periodic in both $\nu$ and $\omega$. Numerics shows that it is in fact independent of $\nu$, $\omega$ and, most importantly, of the other monodromy parameters $\theta_{0,t,*}$. Let us denote this numerical constant by $\hat\Upsilon$ and compute it using a suitable explicit special PV solution.
  
  Consider the tau function
  \beq\label{taupvsp}
  \tau\lb t\rb = t^{\theta_*^2-\frac{1}{8}} e^{\frac{t^2}{32}+\frac{\theta_* t}{2}},
  \eeq
  which satisfies PV with parameters $\theta_0=\theta_t=\frac14$. The monodromy data for this solution are highly non-generic, yet the expansions \eqref{fourier0} and \eqref{fourier8r} around $t=0$ and $t\to +\infty$ remain meaningful. They are described by the limits
  \beq
  \begin{gathered}
  \sigma=\theta_*,\quad\eta\to -i\infty,\quad \omega=0,\quad\xi\to i\infty,\quad X_+\to 0,\quad X_-\to 0.
  \end{gathered}
  \eeq
  While the imaginary axis expansion \eqref{fourier8}  does not make sense for such monodromy, we may still compare the asymptotics of $\tau\lb t\rb$ as $t\to 0$ and $t\to +\infty$. The corresponding connection constant is given by the product $\Upsilon_{0\to +\infty}=\Upsilon_{0\to i\infty}\Upsilon_{i\infty\to +\infty}$, which remains well-defined. It then becomes possible to extract $\hat \Upsilon$ by computing appropriate limits and comparing them with the value of $\Upsilon_{0\to +\infty}$ obtained directly from \eqref{taupvsp}. The final result is given by $\hat \Upsilon=\sqrt{2\pi}\; G^2\lb\tfrac12\rb e^{-\frac{i\pi}{24}}$ and is confirmed by numerics.

  \section{Discussion}
  Let us mention several open research directions related to the present work. First of all, it should be possible to prove rigorously the connection formulas in Conjectures C and D. The relation \eqref{nudefi} between the pairs of asymptotic parameters $\lb\sigma,\eta\rb$  and $\lb \nu,\rho\rb$ can in principle be obtained by the conventional Riemann-Hilbert/WKB techniques such as the one used in \cite{AK} to solve the connection problem between $0$ and $+\infty$. On the other hand, the general framework for computing the connection constants for the tau functions was developed only recently \cite{ILP}. After this work was completed and submitted for publication in March 2018, a paper by Shimomura  \cite{Shi}, which deals with a PV asymptotic problem on the imaginary axis, appeared on Arxiv. It would be interesting to understand its relation to our results.
  
  It is a challenging problem to derive Fredholm determinant and (asymptotic) series representations for the other Painlev\'e equations. Even in the Painlevé V case, one might wonder whether there exist alternative Fredholm determinant representations of $\tau\lb t\rb$, adapted to analysis of the long-distance regime. Such results would be relevant to $\mathcal N=2$ 4D supersymmetric gauge theory, see e.g. \cite{BLMST,BGT,GG}. In particular, one could use them to derive explicit Nekrasov-type series for the confluent $c=1$ CBs of the 2nd kind. They may be also useful in physical applications involving the confluent Heun equation, such as black hole scattering
  \cite{CN,NC} or the Rabi model \cite{CCR}.

  The result that a composition of two chiral vertex
    operators degenerates into an irregular vertex operator intertwining two rank 1
    Whittaker modules may be extended to the $n$-point
    case. Namely, a composition of $n$ chiral vertex operators
    degenerates to an irregular vertex operator intertwining two Whittaker modules of rank $n-1$. As a consequence, the Fourier expansion for the
    Painlev\'e IV tau function conjectured in \cite{Nagoya}
    could be obtained as a limit of the corresponding expansion for  the PVI tau function. 
    Connection constants for the asymptotics of Painlev\'e IV tau function along certain rays at $\infty$ may also be computable from this limit.
  
  It would also be interesting to study irregular fusion transformations, relating e.g. confluent CBs of the 1st and 2nd kind from the above. One may expect the existence of a ``sliding move''
    \ben
         \begin{tikzpicture}[baseline,yshift=-0.3cm,scale=0.8]
         \draw [thick] (-1.2,-0.04) -- (2,-0.04);
         \draw [thick] (-1.2,0.04) -- (2,0.04);
         \draw [thick] (2,0) -- (3,0);
         \draw [thick] (0,0.04) -- (0,1); 
         \draw (-0.7,0) node[above] {\scriptsize $\lb\theta_*,\frac14\rb$};
         \draw (1,0) node[above] {\scriptsize $\lb\frac{\theta_*}{2}-\nu,\frac14\rb$};
         \draw (2.7,0) node[above] {\scriptsize $\theta_0$};
         \draw (0,1) node[right] {\scriptsize $\theta_t$};
         \draw [fill] (2,0) circle (0.08);
         \end{tikzpicture}   
         \lb t\rb \quad = \quad \int_{\mathbb R_+}
      F_{\text{irr}}\left[\substack{\theta_t\vspace{0.08cm} \\ \theta_{*}\;\;\;\theta_0};\substack{\sigma\vspace{0.15cm} \\  \nu}\right] \;\;
         \begin{tikzpicture}[baseline,yshift=-0.3cm,scale=0.8]
         \draw [thick] (-1.2,-0.04) -- (0,-0.04);
         \draw [thick] (-1.2,0.04) -- (0,0.04);
         \draw [thick] (0,0) -- (2,0);
         \draw [thick] (1,0) -- (1,1); 
         \draw (-0.7,0) node[above] {\scriptsize $\lb\theta_*,\frac14\rb$};
         \draw (0.5,0) node[above] {\scriptsize $\sigma$};
         \draw (1.7,0) node[above] {\scriptsize $\theta_0$};
         \draw (1,1) node[right] {\scriptsize $\theta_t$};
         \draw [fill] (0,0) circle (0.08);
         \end{tikzpicture}
   \lb t\rb
   \; d\sigma
      \ebn
 which could provide an analytic meaning to $\mathcal D\lb t\rb$,  so far defined only as a formal series.  The irregular fusion kernel $F_{\text{irr}}$ should in principle be computable by a suitable formal limit of the Ponsot-Teschner formula. This has much to do with the theory of quantum dilogarithms ---  the subject to which Ludvig Faddeev has made fundamental contributions.

       \vspace{0.1cm}
        
          \noindent
          { \small \textbf{Acknowledgements}.  The authors would like to thank B. Carneiro da Cunha, P. Gavrylenko, N. Iorgov, A. Its and T. Oshima for their interest to this work and useful discussions. At the early stages of this project, we have benefited from the comments of A.~Kitaev and M.~Mazzocco. This work was partially supported by JSPS KAKENHI Grant Number JP15K17560, and JSPS and MAEDI under the Japan - France Integrated Action Program (SAKURA).}

 \end{document}